\documentclass[aps,pre,groupedaddress,reprint,showpacs]{revtex4-1}
\usepackage{color}
\usepackage{hyperref}
\hypersetup{colorlinks=true,linkcolor=blue,citecolor=blue,filecolor=blue,urlcolor=blue}
\usepackage{graphicx}
\usepackage{amsmath}
\usepackage{mathptmx}
\usepackage{aas_macros}
\usepackage{enumitem}
\usepackage{multirow,bigstrut,bigdelim}
\begin{document}
\def\my_indent{0.75cm}
%--------------------------------------------------------------------------------------------------
\title{The introductory astronomy course at the University of Cape Town: probing student perspectives}
\author{Vinesh~Rajpaul}
%\email{vinesh.rajpaul@gmail.com}
\affiliation{Academic Development Programme, University of Cape Town, Rondebosch 7701, South Africa}
%\affiliation{Astrophysics, Cosmology and Gravity Centre (ACGC), University of Cape Town, Rondebosch 7701, South Africa}
\thanks{\textcolor{black}{Present address: Sub-department of Astrophysics, Department of Physics, University of Oxford, Oxford OX1 3RH, United Kingdom}}
\author{Saalih~Allie}
\email{saalih.allie@uct.ac.za}
\affiliation{Academic Development Programme, University of Cape Town, Private Bag X3, Rondebosch 7701, South Africa}
\affiliation{Department of Physics, University of Cape Town, Rondebosch 7701, South Africa}
\author{Sarah-Louise Blyth}
\affiliation{Astrophysics, Cosmology and Gravity Centre (ACGC), University of Cape Town, Rondebosch 7701, South Africa}
\date{\today}
%--------------------------------------------------------------------------------------------------
\begin{abstract}

We report on research carried out to improve teaching and student engagement in the introductory astronomy course at the University of Cape Town. This course is taken by a diverse range of students, including many from educationally disadvantaged backgrounds. We describe the development of an instrument, the Introductory Astronomy Questionnaire (IAQ), which we administered as pre- and post-tests to students enrolled in the course. The instrument comprised a small number of questions which probed three areas of interest: student motivation and expectations, astronomy content, and worldview. Amongst our findings were that learning gains were made in several conceptual areas, and that students appeared to develop a more nuanced view of the nature of astronomy. There was some evidence that the course had a positive impact on students' worldviews, particularly their attitudes towards science. We also identified a promising predictor of course success that could in future be used to identify students requiring special teaching intervention.
\end{abstract}
%--------------------------------------------------------------------------------------------------
\maketitle
%--------------------------------------------------------------------------------------------------
\section{Introduction} \label{sec:intro}
%--------------------------------------------------------------------------------------------------
Astronomy is currently a key science focus area in South Africa \citep{whitepaper:1996,rdstrat:2002,Whitelock:2008a}, with the national government having made large investments in facilities over the past ten years in both optical and radio astronomy. For example, the Southern African Large Telescope (SALT) is currently the largest optical telescope in the southern hemisphere, and the upcoming MeerKAT -- a precursor instrument for the Square Kilometre Array (SKA), which will be the world's largest and most sensitive radio telescope, two-thirds of which will be hosted in South Africa -- is to be the largest and most sensitive radio telescope in the hemisphere. In addition to facilities funding, there has also been a strong focus on human capacity development, with a view to increasing the number of South African research astronomers.

The Department of Astronomy within the Faculty of Science at the University of Cape Town (UCT) is, currently, the only one of its kind in South Africa: usually, astronomy courses at other South African universities are taught by physics departments. However, very few physics departments offer astronomy at the undergraduate level \footnote{In South African parlance, the term `Faculty of Science' refers to an organizational structure and not the staff within the structure. The analogous American terminology would be `School of Science' or `College of Science.' `Staff' is usually used in a general sense and includes `faculty' as used in the American sense.}. 

The academic staff members in the Department of Astronomy at UCT are all observational astronomers who are active in research areas from stellar astrophysics to extragalactic studies and observational cosmology. The Department of Astronomy has strong links with the Cosmology and Gravity Group (CAGG) within the Department of Mathematics and Applied Mathematics, whose research themes include string theory, early universe physics, cosmology, and gravitational wave physics. 

As part of the thrust to grow astronomy in South Africa, teaching resources were pooled by a number of universities to form the National Astrophysics and Space Science Programme (NASSP). NASSP, which is currently hosted at UCT, is a graduate programme targeted at South African students (but includes students from other parts of Africa and further afield) pursuing Honours, masters and doctoral degrees\footnote{Compared to bachelors' degrees in the USA, South African degrees tend to be much more rigid. A Bachelor of Science degree in South Africa nominally comprises the equivalent of $9$ full-year courses spread over three years: typically, four first-year courses, three second-year courses and two third-year (major) courses. For example, first-year courses for a student majoring in Astrophysics and Physics could be Physics I, Mathematics I, Applied Mathematics I, Astronomy I (half-course) and Statistics I (half-course). Second-year courses would consist of Astrophysics II, Physics II, and Mathematics II or Applied Mathematics II, while third-year courses would be Astrophysics III and Physics III. The three-year Bachelor of Science degree is followed by a one-year Bachelor of Science (Honours) degree, comparable to the first year of a PhD programme in the USA. Typically, further postgraduate study will entail a research Master of Science degree (nominally two years), followed by a PhD (nominally three years).} under the supervision of astrophysicists and cosmologists from across the country \citep{Whitelock:2008b}. Owing to the history of South Africa, the majority of the population is under-represented in key areas of science and mathematics at all levels, including astronomy. Thus increasing diversity and overall participation rates is a key aim of higher education in South Africa, in particular with regard to the scientific disciplines. At undergraduate level, many special interventions have been put into place, in particular extended degree programmes that aim to bridge the gap between high-school and university. A specific challenge, however, is to increase the proportion of black students at postgraduate level, but there has not been much activity in this regard \footnote{ The terms `black' and `African' are often used interchangeably in the South African context. In this paper, the term `black' is used to include all previously disenfranchised groups, in particular those previously categorized in South Africa as African, Colored, and Indian (of Asian origin).}. However, NASSP has recently initiated a special postgraduate bridging programme at the Honours level, aimed specifically at students from educationally disadvantaged backgrounds, particularly from historically black universities.

Widening participation at undergraduate levels does however lead to student cohorts with broad ranges of backgrounds, abilities and levels of preparation. This poses a challenge to teaching effectively across the spectrum present in a classroom. Therefore it is important that ways of characterizing the heterogeneity present are explored, in order to optimize teaching approaches and to allow all students to participate meaningfully. \textcolor{black}{This applies particularly to the first-year astronomy course at UCT, AST1000F, the focus of this paper.}

\textcolor{black}{In the sections that follow, we describe the AST1000F course in more detail (Section \ref{sec:course}), the development of an instrument which we administered as pre- and post-tests to AST1000F students, in order to probe three different areas of interest (Section \ref{sec:IAQ}), and the methodology used when administering the instrument and analysing the resultant data (Section \ref{sec:methods}). Next, we present our main results (Section \ref{sec:results}), and finally, we discuss our findings and offer conclusions and possibilities for future work (Section \ref{sec:discuss}).}
%--------------------------------------------------------------------------------------------------
\section{The AST1000F course and student cohort} \label{sec:course}
%--------------------------------------------------------------------------------------------------
%---------------------------------------------------------------------------
\subsection{Course context}
%---------------------------------------------------------------------------
At the undergraduate level, the Department of Astronomy at UCT offers students the option of majoring in Astrophysics (usually taken along with a second major in Mathematics, Applied Mathematics, or Physics). In particular, the \emph{Introduction to Astronomy} course, or AST1000F, is a first-year-level astronomy course which has been offered in its current form since 1994 by the Department of Astronomy; broadly it may be thought as an analogue of the U.S.\ ``Astro~101''-type courses discussed in the literature \citep{Partridge:2003,Zeilik:2004}. While AST1000F is \emph{not} a prerequisite for a major in Astrophysics (unlike the the second- and third-year courses astrophysics courses), it is strongly recommended to students intending to major in Astrophysics. AST1000F has no prerequisites, and is open to any student from any faculty within the university. 
%---------------------------------------------------------------------------
\subsection{AST1000F course outline and objectives}
%---------------------------------------------------------------------------
The objectives of AST1000F, from the point of view of the lecturer (S.-L.B.), are:
\begin{itemize}[leftmargin=\my_indent] \itemsep0.5pt
\item to provide an overview of the field of astronomy to undergraduate students from a diverse range of backgrounds;
\item to impress on students how scientific knowledge is gained (i.e.\ how and what data are collected, how data are analyzed, comparison to models/theory); and
\item to expose students to researchers in different fields of astronomy in order to gain more insight into how research is done, as well as to showcase potential role models in science to undergraduate students.
\end{itemize}  

The course content aims to meet the first two objectives, and closely follows the prescribed textbook, \emph{Astronomy Today} by \citet{TEXTBOOK:2010}. The first part of the course covers the Celestial Sphere, the Earth-Moon-Sun system, the history of Astronomy, radiation and spectroscopy, telescopes and detectors. The next large section deals with the Solar System and exoplanets. This is followed by stellar formation and evolution, and the course concludes with topics including galaxies and cosmology. With a view to meeting the third objective, 5 lectures are presented by guests who are experts in different fields of astronomy.

From the student perspective, the intended outcomes of the course are for the student to gain:
\begin{itemize}[leftmargin=\my_indent] \itemsep0.5pt
\item a broader understanding of our place in the universe and the different scales involved, from the size of Earth up to large-scale structures like galaxy superclusters;
\item understanding of and appreciation for the scientific method;
\item knowledge of the terminology and concepts covered in the course material; and
\item insight into how information is gathered and how scientific conclusions are made in the field of astronomy.
\end{itemize}
These outcomes are broadly consistent with those reported for ``Astro~101'' courses in the United States~\cite{Partridge:2003}.

Throughout the course, there was an emphasis of the fact that doing science involves an ongoing process of discovery and change. For example, the lecturer stressed that the current models used to describe astrophysical objects and processes are likely to change on the basis of future observations and theoretical advances.
%---------------------------------------------------------------------------
\subsection{Course format and teaching methods}\label{sec:format}
%---------------------------------------------------------------------------
The UCT academic year (February to November) comprises two semesters. The course runs over the first semester, and comprises 60 lecture periods of $45$~minutes each, and 12 weekly tutorial afternoon sessions of up to 2.5 hours each. Student interaction and discussion are strongly encouraged in lectures. Immediate student feedback is obtained by using in-class multiple choice questions which require students to vote using paper cards.

Occasionally, group tutorials are held in a lecture slot, to assist students to consolidate particular concepts. Material for these tutorials is based on \emph{Lecture-tutorials for introductory astronomy} by \citet{Prather:2008}.

The weekly tutorial sessions vary widely in format. Typically, four of the 12 sessions are tutorial-type sessions where students can ask the lecturer any questions they wish about the course content. Two further sessions involve trips to the local Iziko Planetarium. These sessions are run by the course lecturer and a tutor, and are used to explain celestial coordinates in a three-dimensional setting, as well as to expose students to constellations visible in the night sky. Students also visit the South African Astronomical Observatory (SAAO) where they are able to take part in star-gazing on small telescopes. Three of the twelve tutorial sessions involve exercises in a computer laboratory. These exercises include CLEA tutorials \citep{CLEA:2000} and taking part in citizen cyber-science projects such as \emph{Planet Hunters} \citep{HUNTERS:2011} and the \emph{Galaxy Zoo} \citep{ZOO:2007} projects. The aim of these sessions is to expose students to how scientific data analysis is performed and how information is gathered in astronomy. The CLEA tutorials simulate telescope observations and allow basic analysis using built-in tools of the `data' taken with the simulator to extract useful quantities and illustrate data analysis methods. One practical slot is also used as a poster presentation afternoon (see section \ref{sec:Assessment} for further information on this activity).

In addition to lectures and tutorial sessions, course tutors (TAs) are available for consultation twice per week in the afternoons, when students can informally ask questions and get help with completing problem sets (see below). 

Contact with the students (outside of lectures, tutorials, and consultation sessions with tutours) is maintained using the university online course management system, Vula (based on the Sakai framework). The software enables course-wide emails and maintains a repository of the lecture materials for downloading by the students. 

%---------------------------------------------------------------------------
\subsection{Assessment}\label{sec:Assessment}
%---------------------------------------------------------------------------
Assessment in the course involves a number of inputs. The accumulated class record counts $50\%$ towards the final grade for the course\footnote{Grading of all inputs is done on a numerical rather than alphabetical scale, with $50\%$ constituting a passing grade.}, and the final $2$-hour long examination contributes the remaining $50\%$.

Inputs to the class record include five bi-weekly homework problem sets, the top two out of three class tests, and the marks from a group scientific poster assignment. For the poster assignment the students typically work in groups of four, and design a scientific poster based on some aspect of the Solar System. This allows them to do some deeper research into this area of the course that would not be covered in lectures, as well as exposing them to a real method of research communication which is used in many scientific fields. 

In the class tests (typically $45$~minutes long), assessment ranges from testing knowledge of scientific facts to testing three dimensional visualization of the celestial sphere as well as the magnitude system and the understanding of physical systems such as the Earth-Moon-Sun system. Since the course is open to students from across the university, the mathematics component of the course is kept to a simple level (no more than high school level mathematics is required) and wherever possible, conceptual understanding is tested, rather than students' ability to remember and manipulate formulae.

The final examination is essentially a longer version of the class tests and covers the entire course. The examination also includes a long essay-type question (comprising $15\%$ of the paper) where students choose one out of three possible topics.
%---------------------------------------------------------------------------
\subsection{Students}\label{sec:students}
%---------------------------------------------------------------------------

\begin{table}[h!]
  \centering
  \caption{Summary of demographic information obtained from a sample of 79 students taking the AST1000F course in 2013. Here, ``EGS'' stands for Environmental and Geographic Sciences, and includes such fields as climatology, oceanography, and geology. The numbers presented for (selected) subjects taken in high school refer only to subjects taken by students in their final year of high school.}
    \begin{tabular}{cccc}
    \hline
    \hline
    \multicolumn{2}{c}{Category} & \multicolumn{2}{c}{Count (\%)} \bigstrut\\
    \hline
    \multirow{2}[2]{*}{Gender} & Male & 59 & (75\%) \bigstrut[t]\\
      & Female & 20 & (25\%) \bigstrut[b]\\
    \hline
    \multirow{5}[2]{*}{Nationality} & South African & 55 & (70\%) \bigstrut[t]\\
      & Other African & 8 & (10\%) \\
      & United States & 7 & (9\%) \\
      & Norwegian & 6 & (8\%) \\
      & Other & 3 & (4\%) \bigstrut[b]\\
    \hline
    \multirow{4}[2]{*}{Home language} & English & 54 & (68\%) \bigstrut[t]\\
      & Other South African & 17 & (22\%) \\
      & Norwegian & 6 & (8\%) \\
      & Other & 2 & (3\%) \bigstrut[b]\\
    \hline
    \multirow{4}[2]{*}{Year of study} & 1 & 40 & (51\%) \bigstrut[t]\\
      & 2 & 20 & (25\%) \\
      & 3 & 15 & (19\%) \\
      & 4 & 4 & (5\%) \bigstrut[b]\\
    \hline
    \multirow{5}[2]{*}{Age (years)} & 18 & 21 & (27\%) \bigstrut[t]\\
      & 19 & 21 & (27\%) \\
      & 20 & 13 & (16\%) \\
      & 21 & 13 & (16\%) \\
      & $> 22$ & 11 & (14\%) \bigstrut[b]\\
    \hline
    \multirow{9}[2]{*}{University majors} & Applied mathematics & 12 & (15\%) \bigstrut[t]\\
      & Astrophysics & 14 & (18\%) \\
      & Biology-related major & 9 & (11\%) \\
      & Computer science & 28 & (35\%) \\
      & EGS-related major & 11 & (14\%) \\
      & Engineering & 7 & (9\%) \\
      & Mathematics & 10 & (13\%) \\
      & Other & 5 & (6\%) \\
      & Physics & 17 & (22\%) \bigstrut[b]\\
    \hline
    \multirow{4}[2]{*}{Relevant high-school subjects} & Computer programming & 27 & (34\%) \bigstrut[t]\\
      & Geography & 21 & (27\%) \\
      & Mathematics & 79 & (100\%) \\
      & Physics & 68 & (86\%) \bigstrut[b]\\
    \hline
    \hline
    \end{tabular}%
  \label{tab:demo}%
\end{table}%

The course has grown in student numbers over the years, with an average over the past six years of 115 students. In recent years the course has attracted students mainly from the Faculty of Science, but there has also been representation from other faculties, particularly Engineering and, to a lesser extent, from the Faculty of Humanities. Students who have declared their major as Astrophysics comprise a relatively small fraction of the class (13.6\% on average from 2008-2013).

There are many students who take the course in a year other than their first year of study (something which is relatively unusual for a first-year science course in South Africa) as an elective course, and the course is also popular with semester study abroad students who take it as a transferable credit towards their degrees at their home institutions. Therefore the student cohort is relatively diverse, although the majority of students in recent years have been science majors (in contrast to the situation in the United States, where the majority of students taking an introductory astronomy course are non-science majors \citep{Fraknoi:2002}).

This year, $108$ students signed up for AST1000F, of whom $100$ ended up writing the final examination for the course (the remaining 8 students either dropped the course because of timetable clashes, or were compelled to drop the course on account of not meeting sub-minimum academic criteria for one or more other courses at UCT).

To characterize the 2013 student cohort, we asked the students to complete an optional, online demographics survey. This was done towards the end of the semester, when the class size had already decreased to $100$ students. $79$ students completed the survey; salient data thus obtained are summarized in Table \ref{tab:demo}.

First, it is noteworthy that a third of the class had a home language which was not English -- in fact, 11 different home languages were represented in the class, with 7 of them being South African languages (South Africa has 11 different official languages, and English is only the fourth most-widely spoken home language in the country). Indeed, even though the language of instruction for almost all courses at UCT is English, having a student cohort with a diverse range of home languages is commonplace, and one often speaks of ``language barriers'' in this educational context. Furthermore, in South Africa, having a home language other than English is, for historical reasons, often a good proxy for socio-economic disadvantage \citep{Harber:2001,Utne:2005}. 

We also note that whereas the majority of students ($>85\%$) report having taken physics as part of the Physical Science curriculum in their final year of high school, the South African Physical Sciences curriculum for grades 10-12 (the final three years of school) does not explicitly include any astronomy content. There is, however, some astronomy content in the South African Natural Sciences curriculum for grades 4-9, with particular emphasis placed on the Earth-Sun-Moon system, the solar system, and gravity~\cite{NCSjunior:2002,NCSsenior:2003}.

Finally, the large proportion of students who had computer science as a major (or one of their majors) is anomalous. In other years, there have been significantly more humanities and engineering students, and fewer computer science students, taking the course; the anomaly is attributed to present timetabling logistics.

%--------------------------------------------------------------------------------------------------
\section{Development of a new questionnaire} \label{sec:IAQ}
%--------------------------------------------------------------------------------------------------
%---------------------------------------------------------------------------
\subsection{Motivation for development}\label{sec:motivation}
%---------------------------------------------------------------------------
Physics education research (PER) is well established at the University of Cape Town and has, in part led to various improvements and reforms in physics teaching see, for example, Refs.\ \citep{Allie:1998,Allie:1998b,Campbell:2000,Allie:2003,Buffler:2008,Volkwyn:2008}). Unfortunately, however, these efforts have thus far not included astronomy. Given the increasing interest in astronomy in South Africa, it was felt to be an opportune time to carry out a study that could be used as a starting point to inform aspects of the astronomy curriculum. Thus, of the main motivations for developing and deploying the instrument was to use the data thus obtained to try to optimize the teaching of the AST1000F course, and to maximize meaningful engagement with its diverse cohort of students; in order to do so, as a first step we wanted to know who was taking the course, what motivated them, what they expected to learn, and how and what they thought about astronomy. 

Related to the optimization of astronomy teaching and learning, we also wished to probe issues tied to the fact that a number of students taking the AST1000F course -- as is the case for most courses at UCT -- tend to come from historically-disadvantaged communities and have home languages other than English, which manifests in academic under-performance \citep{Allie:1998b}. Although there are currently bridging and intervention programmes in both astronomy and physics for such students at UCT \citep[e.g.][]{Lubben:2010,Nwosu:2011,Nwosu:2013}, students are often placed into these programmes only after they fail one or more of their courses at UCT. Thus we wished to investigate whether we might be able to use a simple questionnaire to identify potential candidates for such programmes earlier on than is currently done, and perhaps to provide them with extra academic support \emph{ab initio}.

Another primary motivation for developing the questionnaire was to study the ways in which students' views changed (or didn't change) after having taken the course; in particular, we wanted to study the possible ways in which an astronomy course might be used to influence \emph{non-content} issues regarding worldview and scientific thinking \cite{Wallace:2013}, especially given that students who study science at a tertiary level in South Africa have the possibility of ending up in government, policy-formulation or education.
%---------------------------------------------------------------------------
\subsection{Putting together the questionnaire}\label{sec:putting2gether}
%---------------------------------------------------------------------------
In recent years there have been many studies focusing on student understanding with regard to astronomy and cosmology. For example, a number of astronomy-focused instruments and approaches for diagnosing student understanding have been developed ~\citep{2006AEdRv...5a...1H}. This includes instruments such as the Astronomy Diagnostic Test \citep{Hufnagel:2000,Zeilik:2002,Hufnagel:2002,Deming:2002,Brogt:2007}, the Light and Spectroscopy Concept Inventory \citep{Bardar:2006,Bardar:2006b}, the \textcolor{black}{Star Properties Concept Inventory \citep{Bailey:2007,Bailey2012,Lopresto:2009}}, the Lunar Phases Concept Inventory \citep{Lindell:2001,Lindell:2004} and the Newtonian Gravity Concept Inventory \citep{2013AEdRv..12a0107W,Williamson:2012}. In addition student understanding of cosmology was explored in detail in \cite{2011AEdRv..10a0106W,2011AEdRv..10a0107W,2012AEdRv..11a0103W,2012AEdRv..11a0104W,2002AEdRv...1b..28P}. However we were not able to find a suitable (single) instrument that probed the range of issues that were of interest to us in this initial phase of the research. We therefore deemed it necessary to develop our own instrument aimed at our specific needs. We wish to point that many aspects of the afore-mentioned instruments are already used in assessment in the AST1000F course, as noted in Sec.\ \ref{sec:format}; moreover, the aforementioned instruments were significantly too technical for a \emph{pre-course} survey in AST1000F, given most of the students' limited or non-existent astronomy education prior to doing the course (see Sec.\ \ref{sec:students}). Additionally, we wished to tailor the instrument to probe broader issues pertaining to the course or issues specific to the South African context (Secs.\ \ref{sec:students}, \ref{sec:motivation}). 

With regard to the astronomy content, members of the PER research group and members of the Department of Astronomy at UCT used AST1000F curricular materials and AER literature \citep[e.g., in addition to references in the above paragraph,][]{Partridge:2003,Trumper:2001b,Zeilik:2004,Waller:2011} to draw up a preliminary list of about 50 topics, for potential inclusion in the instrument. Concerning non-content issues, the course lecturer (S.-L.B.) noted that she tried to introduce broader aspects of scientific thinking into the curriculum; in previous years, she emphasized the nature and methods of science throughout the course, while the topic of astrology~\cite{2011AEdRv..10a0101S,Caton:2013} was also explicitly addressed during a planetarium practical session.

Taking all of the above into account, along with the constraint that the final instrument would have to be administered during a single lecture period, the following topics for questions were selected: 
\begin{itemize}[leftmargin=\my_indent] \itemsep0.5pt
	\item motivation and expectations regarding the course; 
	\item ideas about what astronomy entails; 
	\item astronomy as a scientific discipline (as opposed to e.g.\ astrology); 
	\item theories and evidence in science;
	\item gravity;
	\item radiation; 
	\item the SKA and radio astronomy; 
	\item sizes/scales in the universe; and
	\item understanding of and ability to explain a few very basic astronomical entities.
\end{itemize}

Whereas a multiple-choice question (MCQ) format may have proven more convenient for analysis, it was decided to use the format modelled on the Physics Measurement Questionnaire \citep{Volkwyn:2005,Volkwyn:2008} in which free-response writing formed the basis of the substantive analysis. Apart from facilitating the expression of views not thought of by us or suggested by the literature, experience with MCQ-formats has highlighted the issue of second-language students misinterpreting questions. For example, even well-established instruments such as the Epistemological Beliefs Assessment for Physical Science \citep{White:1999,Elby:2001} have been shown to be problematic insofar as the interpretation of questions is concerned \cite{Nwosu:2012}. 

The instrument therefore consisted of a number of questions which were framed as a debate between students, followed by a forced choice response (including a ``do not agree with any'' option), and an instruction to give a detailed explanation for the choice made. See Fig.\ \ref{fig:Q7pre} for an example. The instrument also contained a ranking task\cite{2006AEdRv...5a...1H}, an explaining task (paying particular attention to perceived audience \citep{Allie:2008,Allie:2010}), and a free-association task \cite{Johnson:1964,Nelson:2000}. \textcolor{black}{Refinements to the formulation of questions (e.g.\ removing potentially ambiguous or unclear wordings) were made based on informal discussions with a small sample ($\sim15$) of non-astronomy undergraduate students and astronomy graduate students, none of whom was connected to the AST1000F course in any way. For the greater part all the questions appeared to be understood in the manner in which they were intended, and refinements were of a minor nature (for example, ``telescope'' was amended to ``radio telescope'' in the question in Fig.\ \ref{fig:Q7pre}). }

We called the questionnaire the Introductory Astronomy Questionnaire, and shall henceforth refer to it as the IAQ. The full list of questions (without the original formatting) constituting the pre-course IAQ is provided in Appendix \ref{sec:IAQ-pre}.
%-------------------------------------------
% Figure: Sample question from IAQ
%-------------------------------------------
\begin{figure*}[t]
\label{fig:Q7pre}
\centering
\fbox{\includegraphics[width=175mm]{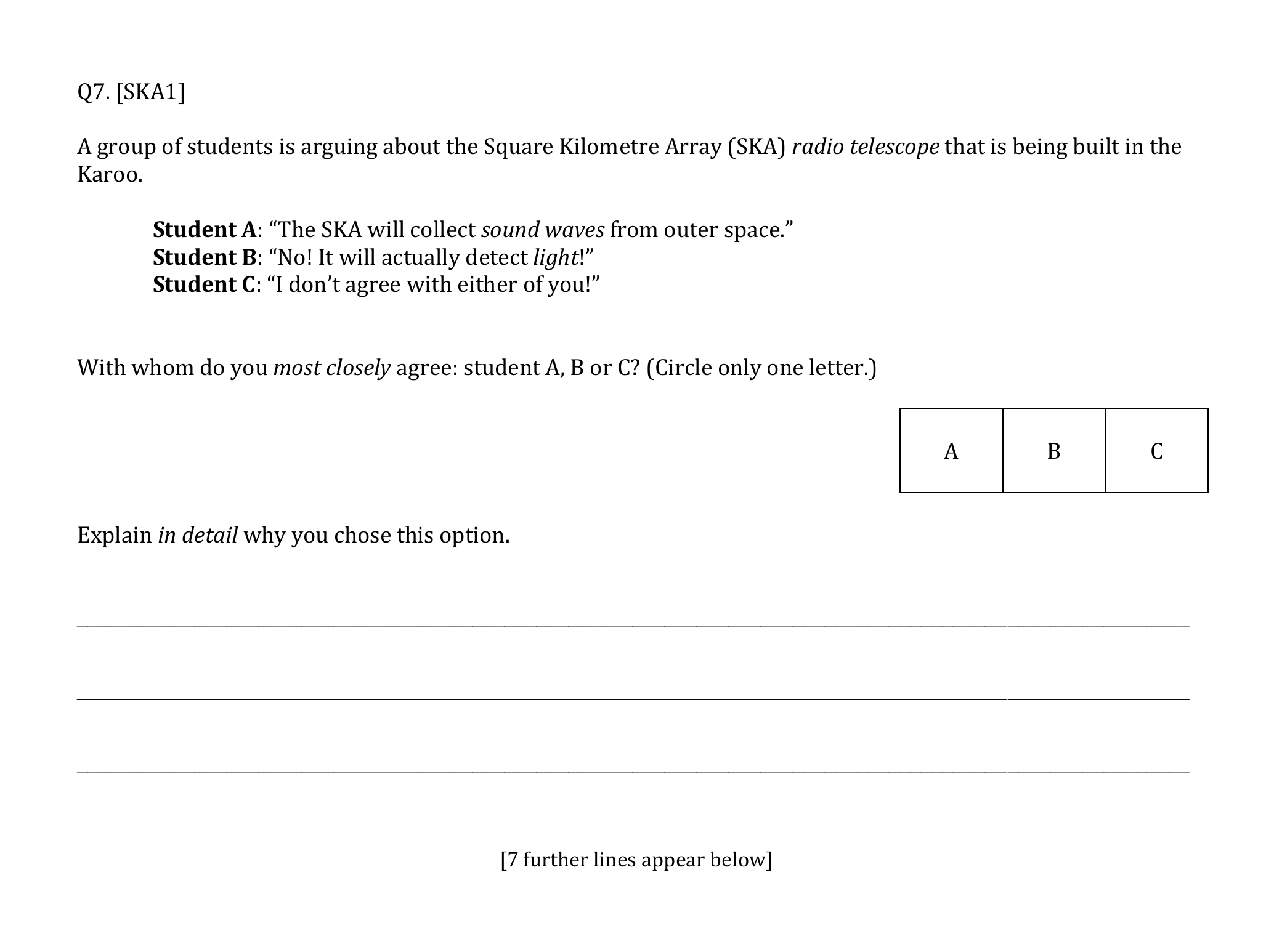}}
\caption{Sample question from the pre-course questionnaire, illustrating the format used for most questions in the instrument. For brevity, a number of the lines provided for students' responses have been truncated. The full list of questions (sans original formatting) may be found in Appendix \ref{sec:IAQ-pre}.\label{fig:Q7pre}}
\end{figure*}
%-------------------------------------------

%---------------------------------------------------------------------------
\subsection{Changes pre-course to post-course}\label{sec:prepost}
%---------------------------------------------------------------------------
Following an analysis of the pre-course responses (see Sec.\ \ref{sec:results}), some changes were made to the post-course version of the IAQ.

The pre-course motivation and expectations questions (Q1a and Q1b in Appendix \ref{sec:IAQ-pre}) were modified to probe post-course experience. Four of the five free-response questions (Q3, Q4, Q5, and Q7 in Appendix \ref{sec:IAQ-pre}) were reduced to multiple-choice questions, where the multiple-choice options were informed by the analysis of the students' pre-course written responses to the same questions. Finally, the radiation question (Q6 in Appendix \ref{sec:IAQ-pre}) was reworded in a less constrained manner in order to allow for a wider range of responses.

The full list of questions constituting the post-course IAQ is provided in Appendix \ref{sec:IAQ-post}.
%--------------------------------------------------------------------------------------------------
\section{Methodology} \label{sec:methods}
%--------------------------------------------------------------------------------------------------

%---------------------------------------------------------------------------
\subsection{Administering the IAQ to students}\label{sec:admin}
%---------------------------------------------------------------------------
The pre-course IAQ was given to students who attended the pre-course orientation lecture in early February 2013. 79 of the 108 students who ended up registering for the course were in attendance (incomplete attendance at such pre-course lectures is typically attributable to the lecture time and venue not yet being known to all students, to some students not yet having finalised their course registration, and/or to some students not knowing they were expected to attend the lecture), and all students in attendance completed the questionnaire.

The IAQ was administered immediately following a short discussion by S.-L.B.\ of some practical, course-related matters (times and venues, forms of assessment, and so on), and importantly, before any discussion or even mention of anything related to the course content.

Students were requested by V.R.\ (who was not otherwise associated with the course in any way) to complete the questionnaire. The following points were made to the students:
\begin{itemize}[leftmargin=\my_indent] \itemsep0.5pt
	\item the questionnaire was to be completed ``as honestly and completely as possible'';
	\item questions were to be answered in the order in which they appeared;
	\item students' answers would be used to improve the course for future cohorts;
	\item there were not necessarily ``right or wrong answers'' for any of the questions;
	\item students' answers would not count in any way toward their grade for the course, and
	\item the course lecturer (S.-L.B.) would not be allowed to match individual answers to student identities, but that V.R.\ or S.A.\ might contact students on an individual-basis \footnote{This served to make students accountable for their answers and not to give whimsical answers; from the responses, it appeared that the instrument was indeed answered seriously.} to clarify what they had written. (For this iteration of the study, at least, we did not end up using the latter option.)
\end{itemize}

The students were given no time limit to answer the questionnaire, although most finished in $20$-$25$~minutes, with none taking longer than about $35$~minutes.

The post-course version of the IAQ was administered in the same manner (both V.R.\ and S.A.\ present) in mid-May 2013, following the last lecture for the course (and a week or two before the final examination for the course). 91 of the $100$ students still registered for the course were in attendance at the lecture, and all filled out the questionnaire. Most finished in around $15$~minutes, with none taking longer than about $30$ minutes.

Of the $91$ students who filled out the post-course IAQ, $71$ had also filled out the pre-course IAQ.
%---------------------------------------------------------------------------
\subsection{Analysis}\label{sec:analysis}
%---------------------------------------------------------------------------
Each student was assigned a unique numerical code which was appended to each page of their IAQ submissions. Submissions were then separated by question, and the analysis was carried out on a question-by-question basis (rather than a student-by-student basis).

For Q1a and Q3--Q7 in the pre-course IAQ (Appendix \ref{sec:IAQ-pre}), and Q6 in the post-course IAQ (Appendix \ref{sec:IAQ-post}), the analysis was carried out using the approach suggested by grounded theory \citep{Corbin:1990,Strauss:1997} as follows. Given the often complex manner in which students expressed themselves, each full written responses was first summarized independently by two researchers, keeping to the original text as closely as possible. Students' answer choices were used, where relevant, to supplement their written responses, although in the occasional cases of a discrepancy between a student's answer choice and their written explanation, precedence was given to the written explanation. (The inter-rater agreement at this stage of the analysis was close to $100\%$.) The summarized form of the students' responses was captured onto a spreadsheet, and used to extract the main points expressed by each student. The inter-rater agreement between the two researchers was over $95\%$ for all questions. 

Using the full list of ``main points'' for each question, a list of fine-grained categories covering one or more main points was drawn up, ensuring that each main point was assigned to a fine-grained category. It should be noted that in most cases, a student's response to a particular question contained more than one main point, and thus ended up being decomposed into more than one fine-grained category. Following an iterative process in which each researcher compared their results and then refined their fine-grained categorization, an average agreement between category assignments of above $90\%$ was obtained.  Again using an iterative process, the fine-grained categories were used to construct broader categories of ideas. These broad emergent categories form the basis of the results presented for these questions.

The analysis for the free-association question (Q2 pre- and post-course) entailed capturing students' responses, correcting them for spelling, grouping equivalent words and phrases together, and then aggregating the results. Further details are provided in Sec.\ \ref{sec:Q2}, along with the results of the analysis.

The analysis for the ranking (Q8a) and explaining (Q8b) tasks entailed capturing all students' responses and identifying incorrect ranking sequences, for the former question, and, for the latter question, capturing responses and assigning scores for the given explanations using an ``incorrect/partially correct/correct'' metric. Further details are provided in Secs.\ \ref{sec:Q8a} and \ref{sec:Q8b}, along with the results.

The analysis for all remaining questions, both pre- and post-course, entailed simply capturing multiple-choice responses.

%--------------------------------------------------------------------------------------------------
\section{Results} \label{sec:results}
%--------------------------------------------------------------------------------------------------
The results from both the pre-course ($79$ students in total) and the post-course ($91$ students) analyses are presented in full so as to characterize the spectrum of responses as broadly as possible. In addition, the change in scores for the ranking and explaining tasks (Secs.\ \ref{sec:Q8a} and \ref{sec:Q8b}, respectively) is also presented for the $71$ students who completed both the pre- and post-course questionnaire.
%---------------------------------------------------------------------------
\subsection{Perspectives on the course itself}
%--------------------------------------------------------------------------- 
%---------------------------------------------------------------------------
\subsubsection{Reasons for doing the course}\label{sec:reasons}
%---------------------------------------------------------------------------
In the pre-course questionnaire, students were asked why they decided to do the AST1000F course. All 79 students answered the question. 

75 out of 79 (95\%) of the responses mentioned as part of the reason for doing the course an \emph{interest} of some sort -- whether an interest in a specific topic, or in astronomy in general -- or a desire to pursue further studies, or indeed a career, in astronomy. 25 out of 79 (32\%) of the responses mentioned a more ``utilitarian'' motivation for choosing the course -- for example, the need of a science elective to fill a gap in an academic curriculum. Finally, 11 out of 79 (14\%) of the responses contained some other reason which did not fit into the two aforesaid categories.

To look more closely at the ``interest''-type motivations (it is worth remarking that most students stated more than one reason for choosing the course, and thus there were more reasons than there were students):
\begin{enumerate}[leftmargin=\my_indent, label*=\Alph*.)] \itemsep0.5pt
	\item 40 students mentioned an interest in astronomy in general, or a field broadly-related to astronomy (physics, space science, etc.). Representative response excerpt: ``Astronomy is a subject that interests me.''
	\item 28 students mentioned an interest in a \emph{specific} astronomy-related topic (e.g.\ galaxies, planets, the origin of the universe). Representative response excerpt: ``I've always been interested in stars, the night sky and theories about the creation and origin of the universe''.
	\item 30 students mentioned a deep or long-standing interest or passion of some sort (an unexpected result). Representative response excerpt: ``It has been my desire since I was young to know more about astronomy.''
	\item 32 students mentioned a general curiosity or desire to broaden horizons as a motivating factor. Representative response excerpt: ``I don't intend to work in astronomy-based field, but would like to learn about the subject.''
	\item 20 students mentioned further studies and/or a possible career in astronomy, physics etc.\ as a motivating factor for taking the course. Representative response excerpt: ``I want to major in astrophysics and this is the only first-year astronomy course offered.''
\end{enumerate}

Of the ``utilitarian''-type motivations, 26 related to credits or gaps in curricula, while only 2 mentioned an impression that AST1000F would be an easy elective course -- effectively quelling concern that the course may have been relatively popular on account of it being perceived as easy rather than interesting.
%---------------------------------------------------------------------------
\subsubsection{Expectations of difficulty}
%---------------------------------------------------------------------------
In the pre-course questionnaire, students were asked how difficult they expected the course was going to be. They were asked  to circle a number from 1 to 5, where 1 meant ``extremely easy'' and 5 meant ``extremely difficult.'' 78 out 79 students answered the question.

\begin{table}[h!]
	\caption{Students' pre-course expectations about the difficulty of the course. Here, 1 means ``extremely easy'' and 5 means ``extremely difficult.''}
	\label{tab:EOD1}
    \begin{tabular}{ccc}
    \hline
    \hline
    Expected difficulty & \multicolumn{2}{c}{Count (\%)} \bigstrut\\
    \hline
    1 & 1 & (1\%) \bigstrut[t]\\
    2 & 12 & (15\%) \\
    3 & 46 & (59\%) \\
    4 & 17 & (22\%) \\
    5 & 2 & (3\%) \bigstrut[b]\\
    \hline
    \hline
    \end{tabular}%
\end{table}

The data obtained from this question (Table \ref{tab:EOD1}) corroborate the findings from the previous question (Sec.\ \ref{sec:reasons}), suggesting that few students chose to do the course in the hope that it would easy.
%---------------------------------------------------------------------------
\subsubsection{Post-course retrospective}
%---------------------------------------------------------------------------
In the post-course questionnaire, students were asked to what extent the topics covered in the course met the expectations they had at the start of the course. They were asked  to circle a number from 1 to 5, where 1 meant ``not at all what [they] expected'' and 5 meant ``exactly what [they] expected.'' Students were also asked how interesting they found the course; again they were asked  to circle a number from 1 to 5, where 1 meant ``not at all interesting'' and 5 meant ``extremely interesting.'' These two questions were answered by 90 out of 91 and 91 out of 91 students, respectively; the results appear in Table \ref{tab:COC1}.

%---------------------------------------------------------------------------
\begin{table}[htb]
	\caption{Students' post-course views on the degree to which their expectations were met vis-\`a-vis the course content, and on how interesting they found the course. Here, 1 means ``not at all what [they] expected'' and ``not at all interesting'', while 5 means ``exactly what [they] expected'' and ``extremely interesting.''}
	\label{tab:COC1}
    \begin{tabular}{cccccc}
		\hline \hline
    Expectations met & \multicolumn{2}{c}{Count (\%)} & Interest in course & \multicolumn{2}{c}{Count (\%)} \bigstrut[b]\\
    \hline
    1 & 0 & (0\%) & 1 & 0 & (0\%) \bigstrut[t]\\
    2 & 3 & (3\%) & 2 & 2 & (2\%) \\
    3 & 14 & (16\%) & 3 & 9 & (10\%) \\
    4 & 37 & (41\%) & 4 & 38 & (42\%) \\
    5 & 36 & (40\%) & 5 & 42 & (46\%) \bigstrut[b]\\
    \hline
    \hline
    \end{tabular}%
\end{table}
%---------------------------------------------------------------------------

It was gratifying to note that the responses to the two questions did not diverge negatively, and that the majority ($>80\%$) of the class had a high level of their expectations met and also found the course interesting  (categories $4$ and $5$ in Table \ref{tab:COC1}).
%---------------------------------------------------------------------------
\subsection{Non-content questions}
%---------------------------------------------------------------------------
%---------------------------------------------------------------------------
\subsubsection{What does astronomy entail?}\label{sec:Q2}
%---------------------------------------------------------------------------
Both pre- and post-course, students were asked to write down the first several words or phrases that came to mind when they were presented with the word ``astronomy''. (This question was asked early in the questionnaire, to avoid prompting the students with specific ideas.) Pre-course, all 79 students polled answered the question, with an average of $10.6$ (standard deviation $4.5$) different words or phrases per student; post-course, all 91 students polled answered the question, with an average of $10.0$ (standard deviation $6.0$) words or phrases per student.

Before aggregating total counts for different words, students' responses were checked for spelling and then processed using a script which grouped equivalent or synonymous words/phrases together: for example, ``Albert Einstein'' and ``Einstein'', ``large'' and ``huge'', ``intriguing'' and ``interesting'',  ``the Earth'' and ``Earth'', and ``star'' and ``stars'' were all deemed to be equivalent. In total, approximately 100 possible pairs of such equivalents were identified.  On the other hand, words which were not necessarily synonymous were \emph{not} grouped together: for example, ``sun'' and ``star'', or ``planet'' and \hbox{``exoplanet''}. Following this processing, a total of $834$ separate words/phrases in the pre-course dataset was reduced to $244$ unique words/phrases, and a total of $913$ words/phrases in the post-course dataset was reduced to $250$ unique words/phrases.

%---------------------------------------------------------------------------
\begin{table}
	\caption{Words/phrases students associated with astronomy, both pre- and post-course (only the words mentioned by at least $10\%$ of the class are shown). In each case, the words are listed by the total number of students whose responses contained the word (or an equivalent). Words with asterisks appear in only one column, i.e.\ either pre- or post-course only.}
	\label{tab:Q2prepost}
\begin{tabular}{cccccc}
\hline
\hline
\multicolumn{ 3}{c}{Pre-course ($n=79$)} & \multicolumn{ 3}{c}{Post-course ($n=91$)} \\
\hline
Word/phrase & \multicolumn{ 2}{c}{Count (\%)} & Word/phrase & \multicolumn{ 2}{c}{Count (\%)} \\
\hline
     Stars &         74 &     (94\%) &      Stars &         71 &     (78\%) \\

   Planets &         55 &     (70\%) &   Galaxies &         55 &     (60\%) \\

  Galaxies &         53 &     (67\%) &   Universe &         50 &     (55\%) \\

  Universe &         37 &     (47\%) &    Planets &         49 &     (54\%) \\

Black holes &         31 &     (39\%) & Telescopes &         30 &     (33\%) \\

     Space &         30 &     (38\%) & Black holes &         27 &     (30\%) \\

Telescopes &         25 &     (32\%) &      Space &         24 &     (26\%) \\

      Moon &         18 &     (23\%) & Dark matter &         22 &     (24\%) \\

       Sun &         16 &     (20\%) & Supernovae &         22 &     (24\%) \\

Constellations * &         14 &     (18\%) &   Big Bang &         16 &     (18\%) \\

 Milky Way * &         13 &     (16\%) & Solar system &         16 &     (18\%) \\

Solar system &         11 &     (14\%) &       Moon &         14 &     (15\%) \\

  Big Bang &         11 &     (14\%) &        Sun &         13 &     (14\%) \\

Dark matter &         10 &     (13\%) &      Light * &         13 &     (14\%) \\

Supernovae &         10 &     (13\%) & Spectroscopy * &         13 &     (14\%) \\

   Physics &         10 &     (13\%) &    Gravity * &         10 &     (11\%) \\

    Comets * &          9 &     (11\%) & Dark energy * &         10 &     (11\%) \\

   Nebulae * &          9 &     (11\%) &    Physics &          9 &     (10\%) \\

     Earth * &          8 &     (10\%) & Interesting * &          9 &     (10\%) \\

      Time * &          8 &     (10\%) &  Radiation * &          9 &     (10\%) \\
\hline
\hline
\end{tabular}    
\end{table}
%---------------------------------------------------------------------------

%---------------------------------------------------------------------------
\begin{table}
	\caption{Top 15 words/phrases which students associated with astronomy more prominently post-course than pre-course. The words are ranked by improvements in rank from pre-course to post-course; words with fewer than 5 occurrences both pre- and post-course are excluded from the list.}
	\label{tab:Q2delta}
\begin{tabular}{ccccccc}
\hline
\hline
           &     \multicolumn{ 3}{c}{Post-course ($n=91$)} &      \multicolumn{ 3}{c}{Pre-course ($n=79$)} \\
					\hline

Word/phrase & \multicolumn{ 2}{c}{Count (\%)} &       Rank & \multicolumn{ 2}{c}{Count (\%)} &       Rank \\
\hline
Spectroscopy &         13 &     (14\%) &         11 &          1 &      (1\%) &         22 \\

Dark energy &         10 &     (11\%) &         12 &          1 &      (1\%) &         22 \\

 Radiation &          9 &     (10\%) &         13 &          0 &      (0\%) &          - \\

 Cosmology &          8 &      (9\%) &         14 &          1 &      (1\%) &         22 \\

Interesting &          9 &     (10\%) &         13 &          3 &      (4\%) &         20 \\

     Light &         13 &     (14\%) &         11 &          5 &      (6\%) &         18 \\

   Quasars &          6 &      (7\%) &         16 &          0 &      (0\%) &          - \\

HR diagram &          5 &      (5\%) &         17 &          0 &      (0\%) &          - \\

   Gravity &         10 &     (11\%) &         12 &          5 &      (6\%) &         18 \\

Extraterrestrial &          5 &      (5\%) &         17 &          0 &      (0\%) &          - \\

Dark matter &         22 &     (24\%) &          8 &         10 &     (13\%) &         13 \\

Supernovae &         22 &     (24\%) &          8 &         10 &     (13\%) &         13 \\

      Life &          6 &      (7\%) &         16 &          2 &      (3\%) &         21 \\

Exoplanets &          6 &      (7\%) &         16 &          2 &      (3\%) &         21 \\

    Hubble &          6 &      (7\%) &         16 &          3 &      (4\%) &         20 \\
\hline
\hline
\end{tabular}  
\end{table}
%---------------------------------------------------------------------------

In {Table~\ref{tab:Q2prepost}} we present the most prevalent words/phrases identified in students' responses, both in the pre- and post-course datasets. Only the words mentioned by at least $10\%$ of the class are shown. Both the pre- and the post-course lists could be argued to form a reasonable sketch of what astronomy entails. It is interesting to note that the same $7$ words were used by more than a quarter of the class both pre- and post-course. This could have implications for the future marketing of the course!

To home in on the impact of the course on students' ideas about what astronomy entails, we present in {Table~\ref{tab:Q2delta}} the top 15 words/phrases which featured \emph{more prominently} in the post-course dataset than the pre-course dataset. We used the change in the \emph{rank} of a word's prevalence, rather than the change in the proportion of students who cited the word, as a more robust indicator of a change in prevalence \footnote{For example, whereas words such as ``stars'', ``galaxies'', and ``planets'' all saw drops in the proportion of students citing them, these words were very popular ($>50\%$) both pre- and post-course, and indeed their ranks remained essentially unchanged.}. We exclude from the list all words with fewer than 5 occurrences both pre- and post-course.

The words which feature more prominently post-course than pre-course (``spectroscopy'', ``dark energy'', ``radiation'', and so on) are largely more technical concepts which were taught in the course, suggesting a shift from a popular view of astronomy to a more technical one.

Interestingly, the only word to drop more than 3 ranks (and with more than 5 occurrences either pre- or post-course) was ``constellations'': 14 instances pre-course, and 5 post-course, for a drop of 7 ranks. Again, this was consistent with the treatment of the topic in the course (while the topic was presented, it received little emphasis).
%---------------------------------------------------------------------------
\subsubsection{Astronomy versus astrology}
%---------------------------------------------------------------------------
In the pre-course questionnaire (Q3), the question involving astrology sought to prompt students both to explain the difference between astronomy and astrology, and to say whether they thought each one was useful (and why, or why not).

72 students attempted to define, or to differentiate between, astronomy and astrology. Of these 72 students:
\begin{enumerate}[leftmargin=\my_indent, label*=\Alph*.)] \itemsep0.5pt
	\item 47 (65\%) gave explanations which could be reconciled with standard definitions or characterizations of astronomy and astrology. Representative response excerpt: ``Whilst astronomy is a scientific study of the stars, astrology is the theory that the stars help shape each person's destiny and character.'' 
	\item 25 (35\%) gave incorrect (``I am pretty sure they are both the same thing''), vague, or non-committal (``Astronomy and astrology are two different things, but they also have things in common'') explanations.
\end{enumerate}

54 students commented on the usefulness of astrology. Of these 54 students:
\begin{enumerate}[leftmargin=\my_indent, label*=\Alph*.)] \itemsep0.5pt
	\item 13 (24\%) said they thought astrology was useful, or useful if one chose to believe in it. Representative response excerpt: ``A deeper look into astrology proves very interesting and, strangely enough, accurate. Therefore it, too, has its uses.''
	\item The remaining 41 students (76\%) disputed the usefulness of astrology. Representative example: ``The only real use for astrology is to give the masses something mythical to believe in.'' Furthermore, at least 18 (33\%) giving a strongly-worded or scathing critique of astrology (``[Astrology] is total pseudo-science and a load of hogwash if you ask me.'')
\end{enumerate}
Only 25 students commented specifically on the usefulness of astronomy: all said that they thought it was useful, although only $13$ of them provided justification for this statement (factors mentioned included allowing us to learn more about the universe, drive technological development, expand civilization beyond the Earth, and inform predictions about space weather). \textcolor{black}{This is broadly consistent with the findings of \citet{Wallace:2013}, who found that students enrolled in the general education, introductory astronomy ``Astro 101 mega-course'' at the University of Arizona (a large public university in the USA) generally had positive or at least not negative views about the relationship between science and many other aspects of society.}

As noted in Sec.\ \ref{sec:prepost}, the post-course version of the question was simpler, and asked students to choose between one of three distinct options (corresponding to different positions in a hypothetical argument). 91 students answered the question:
\begin{enumerate}[leftmargin=\my_indent, label*=\Alph*.)] \itemsep0.5pt
	\item 81 (89\%) took the position that ``astronomy is a useful scientific discipline, whereas there is no evidence to support any of the claims made by astrology, which is mere superstition.'' 
	\item 3 (3\%) took the position that ``there is plenty of evidence to support astrology, and it can be just as useful as astronomy.''
	\item 7 (8\%) took the position that ``ultimately, astronomy and astrology are both belief systems. Either one of them can be useful, if you choose to believe in it.''
\end{enumerate}

The decrease in the proportion of students indicating that they thought astrology to be useful is consistent with the topic being addressed directly during the course (as suggested in Sec.\ \ref{sec:putting2gether}). Firstly, during a section focusing on the history of astronomy, the lecturer commented on the historical role of the astrology, and also voiced her opinions on astrology (negative), but no class discussion took place on the subject. Secondly, there was a particular effort made to ``debunk'' astrology during a practical planetarium session, as follows. The course lecturer illustrated the effect of precession of the earth's orbit by dialling the star positions back 3000 years, without the knowledge of the students, and projecting the zodiac constellations and the ecliptic onto the planetarium dome. For this epoch ($\sim3000$~B.C.E.), the constellations of the zodiac coincided with the dates on the ecliptic used in conventional horoscopes, i.e.\ the dates purported to be associated with particular star signs. She read the students horoscope prediction for particular star signs (i.e.\ sets of dates) from a magazine, and discussed whether they thought this might apply to them. She then revealed to the class that the epoch setting was in fact $\sim3000$ years in the past. By precessing the celestial sphere to the year 2013, she illustrated the shift of the zodiac constellations with respect to the ecliptic, and then read them the ``newly-applicable'' horoscopes from the same magazine. The students seemed to enjoy the exercise and there was much laughter during this session.

\textcolor{black}{It is interesting to compare these findings with those of \citet{2011AEdRv..10a0101S}, who found that more than three quarters of a sample of nearly 10,000 undergraduates at the University of Arizona considered astrology to be very or ``sort of'' scientific. The authors suggested, however, that while science classes could counter acceptance of pseudoscience, students' scientific literacy need not correlate positively with an understanding that astrology is pseudoscientific; counter-intuitively, broader acceptance of pseudoscience often seems to coexist with strong performance on science knowledge indicators. They further suggest that instructional strategies designed to combat beliefs in astrology may not, therefore, have a larger benefit.}
%---------------------------------------------------------------------------
\subsubsection{The Big Bang as a theory}
%---------------------------------------------------------------------------
In the pre-course questionnaire, the question involving the Big Bang sought to prompt students to comment on their beliefs regarding the Big Bang theory, and in particular, on their understanding of the word ``theory'' in this context. 78 students answered the question.

76 students alluded to their belief in whether the Big Bang happened. Of these students:
\begin{enumerate}[leftmargin=\my_indent, label*=\Alph*.)] \itemsep0.5pt
	\item 34 (45\%) indicated that they believed the Big Bang did or likely did take place. Representative response excerpt: ``The Big Bang is a theory but I believe it is indeed how the universe was created.''
	\item 26 (34\%) adopted an agnostic position, saying either that they weren't sure, or they thought it unknown or unknowable whether the Big Bang really did take place. Representative response excerpt: ``I don't think one can say with absolute certainty that the universe started with the Big Bang. No one actually knows.''
	\item 16 (21\%) said they did \emph{not} believe the Big Bang took place, or that it was unlikely. Representative response excerpt: ``I don't believe that the universe started with the Big Bang -- I believe that God created the universe.''
\end{enumerate}
Of those in the first two categories, i.e.\ those who did not reject the possibility of the Big Bang taking place, 7 (12\%) said they could reconcile this view with their religious/theistic convictions; on the other hand, of those who said they did not believe the Big Bang took place, 7 (44\%) said they rejected it on religious grounds.

54 students commented specifically on their interpretation of the word ``theory''. Of these students:
\begin{enumerate}[leftmargin=\my_indent, label*=\Alph*.)] \itemsep0.5pt
	\item 32 (59\%) made reference to the Big Bang being a \emph{scientific theory} that is well-tested, falsifiable, supported by observational evidence, and/or widely-accepted in the scientific community. Representative response excerpt: ``All current physical and theoretical evidence, to my knowledge, confirms [that the universe started with the Big Bang]. One example would be the cosmic microwave background radiation.''
	\item 22 (41\%) suggested that the Big Bang was more synonymous with unsubstantiated speculation, or that it was ``nothing more than'' a theory. Representative response excerpt: ``There is no real evidence of the Big Bang occurring because of it [allegedly] happening billions of years before mankind.''
\end{enumerate}

As noted in Sec.\ \ref{sec:prepost}, the post-course version of the question was simpler, and asked students to choose between one of four distinct options (corresponding to different positions in a hypothetical argument). 89 students answered the question:	
\begin{enumerate}[leftmargin=\my_indent, label*=\Alph*.)] \itemsep0.5pt
	\item 42 (47\%) took the position that ``there is strong evidence that the universe started with the Big Bang.''
	\item 21 (24\%) took the position that ``while there may be evidence for the Big Bang, I am not convinced that is how the universe started.''
	\item 25 (28\%) took the position that ``whether or not the Big Bang happened, I believe that God created the universe.''
	\item 1 (1\%) took the position that ``the Big Bang is nothing more than a theory. There is no real evidence that it happened.'' 
\end{enumerate}

The notable change from pre-course to post-course is the decrease in the proportion of students who dismissed the Big Bang theory as being akin to unsubstantiated speculation. The fairly large proportion of students who adopted an agnostic position post-course may be interpreted as an indicator of healthy scepticism, or an understanding that a scientific theory can be tested and falsified but never proven. Finally, the increase in the proportion of students expressing their theistic convictions is, possibly, a result of their feeling more comfortable choosing this option when presented it in multiple-choice format, rather than having to offer it of their own volition. It is also possible that the ethos of the course allowed students to reconcile their religious views with scientific thinking.

\textcolor{black}{For comparison, \citet{Trouille:2013} found that a significant fraction of students enrolled in an undergraduate, general education astronomy course at the Chicago State University (a minority serving institution in the USA) entered the course with alternative conceptions about the Big Bang theory. The alternative conceptions included that there is no evidence in support of the Big Bang theory, that the theory describes the creation of planets, that the theory refers to an explosion within a small point or mass, and that the universe always existed. Following a semester of instruction, the authors found that student understanding of the early universe was more robust, but noted that some mental models about the universe beginning with an explosion seemed difficult to replace.}

\textcolor{black}{In a similar vein, \citet{Prather:2002} found that more than two-thirds of a sample of non-science majors at the University of Arizona thought of the Big Bang theory as describing an ``explosion of pre-existing matter'' prior to instruction. \citet{2012AEdRv..11a0104W} found that more than half of undergraduate students drawn from thirteen different higher-education institutions in the USA referred to the Big Bang theory as describing an explosion, with about an additional third of them speaking of matter existing before the Big Bang.}
%---------------------------------------------------------------------------
\subsection{Scientific-knowledge questions}
%---------------------------------------------------------------------------
Note that the pre- and post-course questions on radiation (Q6 in Appendices \ref{sec:IAQ-pre} and \ref{sec:IAQ-post}) will not be discussed in the present paper as they form part of a separate study, and will be reported on elsewhere.
%---------------------------------------------------------------------------
\subsubsection{Gravitation}
%---------------------------------------------------------------------------
In the pre-course questionnaire, the question involving the pen being dropped on the moon sought to probe students' conceptions about the gravitational force and its universality.

Pre-course, all $79$ students answered the question. Of these students:
\begin{enumerate}[leftmargin=\my_indent, label*=\Alph*.)] \itemsep0.5pt
	\item 53 (67\%) said that the moon had its own gravitational field (albeit weaker than the Earth's), or characterized gravitation in a manner consistent with Newton's law of universal gravitation, and that the pen would therefore fall to the moon's surface. Representative response excerpt: ``Every massive body in the universe has a degree of gravitational attraction to other masses. Therefore the pen would fall onto the surface of moon, but would take longer than on Earth.''
	\item 12 (15\%) said that the pen would \emph{not} float forever because of the presence of the Earth's gravity. Representative response excerpt: ``The force of gravity is always directed towards the Earth. Everything that is dropped from any point above the Earth will fall to the Earth regardless of how far it is above the Earth.''
	\item 9 (11\%) said that the pen would float forever, either because the moon's gravity is too weak and/or the pen is too light, or because there is no gravity at all in space or on the moon. Representative response excerpt: ``There is no gravity on the moon to pull the pen down, and thus the pen will float forever''
	\item 5 (6\%) said that they did not know, or gave a non-committal answer. Representative response excerpt: ``I'm not completely sure about the rules governing gravity.''
\end{enumerate}

As noted in Sec.\ \ref{sec:prepost}, the post-course version of the question was simpler, and asked students to choose between one of three distinct options (corresponding to different positions in a hypothetical argument). 91 students answered the question:
\begin{enumerate}[leftmargin=\my_indent, label*=\Alph*.)] \itemsep0.5pt
	\item 69 (76\%) took the position that ``if you drop a pen on the moon, it will fall to the surface of the moon.'' 
	\item 17 (19\%) took the position that ``the pen will not fall down, it will simply float.''
	\item 5 (5\%) took the position that ``the pen will drift away from the moon and fall towards the Earth.''
\end{enumerate}

The proportion of students giving scientifically-compatible responses, both pre- and post-course, is \textcolor{black}{better than reported elsewhere (see e.g.\ \citet{Noce:1988} and \citet{2013AEdRv..12a0107W} for results from related investigations in Italy and the USA, respectively)}; it is possible that this is due to the fact that Newtonian gravitation is included in the standard South African high school physics curriculum \cite{NCSjunior:2002,NCSsenior:2003}, but this point requires further investigation. It is also interesting to note that a comparable proportion of students gave incorrect answers pre-course ($26\%$) and post-course ($24\%$). This could be attributed to the fact that Newtonian gravity is not a strongly-emphasized topic in AST1000F. While Newton's Law of Universal Gravitation is explained conceptually, it is not assessed directly in tests or assignments; it is, however, used to explain the underlying physics of Kepler's Laws, and is invoked often throughout the course, particularly to demonstrate how the masses of astronomical bodies can be measured by observing their orbits. The original motivation for following this approach was that the majority of the class, being science students, would have taken physics at high school. However, in view of these findings, it may be beneficial to rethink this approach.

Interestingly, informal discussions with about ten senior astronomy undergraduates revealed several alternative conceptions related to gravity. For example, asking students to explain why the planets orbited the sun (rather than falling into the sun), or why astronauts appeared to float in the International Space Station, elicited many incorrect explanations. It is clear that the topic of gravity is worth pursuing in more detail at all undergraduate \textcolor{black}{levels \citep{Noce:1988,Galili:1996,Williamson:2012}.}

%---------------------------------------------------------------------------
\subsubsection{The SKA and radio telescopes}
%---------------------------------------------------------------------------

In the pre-course questionnaire, the question involving the SKA sought to probe students' ideas about what radio telescope do.

Pre-course, $73$ students answered the question. Of these students:
\begin{enumerate}[leftmargin=\my_indent, label*=\Alph*.)] \itemsep0.5pt
	\item 16 (22\%) said that the SKA will collect radio waves, electromagnetic waves, or electromagnetic radiation. Representative response excerpt: ``The SKA will collect \emph{radio} waves, a form of electromagnetic radiation with a very long wavelength, [and] definitely not visible light. It may have been visible light once, but has since been stretched to become radio waves.'' Some students in this category mentioned that electromagnetic radiation is sometimes referred to, loosely, as ``light''.
	\item 17 (23\%) said that the SKA will collect, detect or amplify \emph{light}, with no mention of radio waves or anything to distinguish visible light from longer-wavelength radiation. Representative response excerpt: ``I think it will detect light, because we need light to see. Therefore, the telescope detects light in order for us to see stars etc.''
	\item 19 (26\%) said that the SKA will collect or detect sound. Representative response excerpt: ``I believe a radio telescope would collect sound waves and produce a digital picture by emitting waves and monitoring the time taken for them to return.''
	\item 9 (12\%) said that the SKA will collect sound \emph{and} either light or electromagnetic waves of some form. Representative response excerpt: ``The telescope will collect both sound and light waves from outer space and will gather all this data and process it.''
	\item 12 (16\%) said that they weren't sure know, or that they'd never heard of the SKA. Representative response excerpt: ``I don't know anything about the SKA.''
\end{enumerate}

As noted in Sec.\ \ref{sec:prepost}, the post-course version of the question was simpler, and asked students to choose between one of three distinct options (corresponding to different positions in a hypothetical argument). 91 students answered the question:
\begin{enumerate}[leftmargin=\my_indent, label*=\Alph*.)] \itemsep0.5pt
	\item 76 (84\%) took the position that ``radio telescopes collect a particular type of electromagnetic wave.''
	\item 7 (8\%) took the position that ``like all telescopes, it will collect and amplify light that is too faint for our eyes to see.'' 
	\item 8 (9\%) took the position that ``radio telescopes collect sound waves from outer space.''
\end{enumerate}

Two of the key sections in the course were an introduction to the electromagnetic spectrum and spectroscopy. The terms ``light'' and ``EM radiation'' were used interchangeably throughout the course, following the example of the course textbook~\cite{TEXTBOOK:2010}. The difference between sound waves and electromagnetic waves was stressed, and the particle-wave dual nature of light was discussed. In addition, another key module of the course focused on telescopes and their designs to enable observations of EM radiation across the spectrum. Radio telescopes and, in particular, the Square Kilometre Array (SKA) and South Africa's role in it, received substantial attention. This is, perhaps, the reason for the reduction in responses in the post-course answers referring to the SKA collecting sound waves, since this misconception was directly addressed in the course material.

\textcolor{black}{While instruments do exist which seek to probe specific aspects of students' understanding of radio waves -- most notably the Light and Spectroscopy Concept Inventory \citep{Bardar:2006,Bardar:2006b} -- it seems that there is to be no data available in the literature against which to make a direct comparison of our findings. For example, \citet{Bardar:2007} showed that introductory astronomy students at a number of colleges in the USA entered courses with a poor understanding of the differences (in terms of energy and propagation speed) between visible light and radio waves, and that this understanding improved significantly following instruction. There are also discussions in the literature of alternative conceptions related to optical telescopes \citep{plait:2002,marx:2004}. However, none of this work seems to shed light on students' ideas about \emph{radio} telescopes.} 

%---------------------------------------------------------------------------
\subsubsection{Ranking task}\label{sec:Q8a}
%---------------------------------------------------------------------------
%---------------------------------------------------------------------------
\begin{table*}[htb!]
	\caption{Students' errors on the ranking task, both pre- and post-course. The leftmost column indicates errors made (e.g.\ the first entry corresponds to planets being bigger than stars), and the next columns give the numbers and percentages of students who made these errors.}
	\label{tab:Q8rank}
    \begin{tabular}{ccccccccc}
    \hline
    \hline
     & \multicolumn{4}{c}{Full sample} & \multicolumn{4}{c}{Matched students} \bigstrut\\
    \hline
      \multirow{2}{*}{Incorrect ranking} & \multicolumn{2}{c}{Pre-course} & \multicolumn{2}{c}{Post-course} & \multicolumn{2}{c}{Pre-course} & \multicolumn{2}{c}{Post-course} \bigstrut[t]\\
      & \multicolumn{2}{c}{($n=79$)} & \multicolumn{2}{c}{($n=91$)} & \multicolumn{2}{c}{($n=71$)} & \multicolumn{2}{c}{($n=71$)} \bigstrut[b]\\
    \hline
    Planet $>$ star & 11 & (14\%) & 3 & (3\%) & 7 & (10\%) & 1 & (1\%) \bigstrut[t]\\
    Planet $>$ solar system & 3 & (4\%) & 0 & (0\%) & 1 & (1\%) & 0 & (0\%) \\
    Planet $>$ galaxy & 3 & (4\%) & 0 & (0\%) & 1 & (1\%) & 0 & (0\%) \\
    Star $>$ solar system & 3 & (4\%) & 0 & (0\%) & 1 & (1\%) & 0 & (0\%) \\
    Star $>$ galaxy & 1 & (1\%) & 0 & (0\%) & 1 & (1\%) & 1 & (1\%) \\
    Solar system $>$ galaxy & 7 & (9\%) & 2 & (2\%) & 7 & (10\%) & 0 & (0\%) \\
    Galaxy $>$ universe & 1 & (1\%) & 0 & (0\%) & 0 & (0\%) & 0 & (0\%) \bigstrut[b]\\
    \hline
    \hline
    \end{tabular}%
\end{table*}
%---------------------------------------------------------------------------

Students were asked to rank the following from smallest to largest: galaxy, planet, star, universe, solar system.

Pre-course, 60 out of 79 students ($76\%$) made no mistakes, i.e.\ \textcolor{black}{obtained} perfect scores, on this ranking task; post-course, 86 out of 91 students ($95\%$) made no mistakes on the same task. Looking at the sample of $71$ students who answered the question both pre- and post-course, $57$ ($80\%$) made no mistakes pre-course while $69$ ($97\%$) made no mistakes post-course. The mistakes students made are detailed in Table \ref{tab:Q8rank}. 

\textcolor{black}{Students' improved performance on this ranking task was consistent with the findings of \citet{Coble:2013}, who studied Chicago State University students' understanding of distances and structure in the Universe. While they do advocate explicitly teaching the hierarchical nature of structure in the universe, the aforesaid authors caution that students seem to find ranking tasks easier than tasks dealing with absolute scales, and that they might not be consistent in their reasoning between the two types of tasks.}
%---------------------------------------------------------------------------
\subsubsection{Explaining tasks}\label{sec:Q8b}
%---------------------------------------------------------------------------
Both pre- and post-course, students were asked to provide brief explanations (for their ``friend's 12-year old sister'') of the following: galaxy, planet, star, universe, solar system. Note that these were the same entities students were asked to sort in the ranking task.

Given that the students were encouraged to provide brief and non-technical explanations -- the idea being to probe whether they had a qualitatively correct understanding of the entity in question, and one which they could communicate to someone else, rather than whether they could produce a detailed technical explanation -- the students' answers were marked according to the following scheme:

\begin{itemize}[leftmargin=\my_indent] \itemsep0.5pt
	\item 0 points were awarded for an incorrect explanation;
	\item half a point was awarded for a partially correct (``halfway-there'', or ``on the right track'') explanation; and
	\item 1 point was awarded for a minimally-correct (``adequate'') explanation, as detailed below.
\end{itemize}
Criteria for an explanation to qualify as minimally-correct ($1.0$~points) were fixed as below.
\begin{itemize}[leftmargin=\my_indent] \itemsep0.5pt
	\item \textbf{Galaxy} -- student indicated that it is a collection/system of stars and other material, and provided any information to distinguish it from e.g.\ a stellar system or star cluster (e.g.\ student mentions ``billions of stars'').
	\item \textbf{Planet} -- student indicated that it is an object in orbit around the sun (or another star), and provided any information to distinguish it from e.g.\ an asteroid or comet (larger than a certain size, stable due to its own gravity, cleared its immediate neighbourhood, etc.).
	\item \textbf{Star} -- student indicated that it is a large/massive, hot/luminous sphere of plasma/ball of gas, or any equivalent explanation.
	\item \textbf{Universe} -- student indicated that it is all existing matter and space, all of the cosmos, everything, the totality of existence, a connected space-time, or any equivalent explanation.
	\item \textbf{Solar system} -- student indicated that it is the sun and the objects in orbit around it (e.g.\ planets, moons), \emph{or} that it is a system comprising one or a small number of stars which orbit each other. 
\end{itemize}
Students giving an explanation which only partially matched the criteria set out above for a given object were awarded $0.5$~points (e.g.\ ``a planet is a body which orbits the sun), as were students who provided examples without further explanation (e.g.\ ``a star is something like the sun''). Students not writing anything which matched the above criteria, and/or giving any factually-incorrect information, were awarded $0.0$~points (e.g.\ ``a planet is any place that can support life'').

Students' explanations were marked independently by V.R.\ and S.-L.B.; agreement in the scores assigned was found to be $>95\%$ for all objects. The scores are tabulated in Table \ref{tab:Q8exp} for the sample of matched students ($n=71$) who answered the question both pre- and post-course (see Appendix \ref{sec:exp-full} for the results for the full pre- and post-course samples). For each object, the total number of students answering the question, the number of students scoring $0.0$, $0.5$ or $1.0$ out of $1.0$, and the mean score for the object, is shown; blank responses were not assigned scores. Note that many students interpreted ``solar system'' as ``stellar system''; scores for the two different possible interpretations are shown separately.

%---------------------------------------------------------------------------
\begin{table*}[tbh!]
	\caption{Scores on the explaining task, for the sample of matched students ($n=71$) who answered the question both pre- and post-course.}
	\label{tab:Q8exp}
    \begin{tabular}{ccccccccccc}
    \hline
    \hline
      & \multicolumn{5}{c}{Pre-course ($n=71$)} & \multicolumn{5}{c}{Post-course ($n=71$)} \bigstrut\\
    \hline
    Object & $n(\textrm{total})$ & $n(1.0)$ & $n(0.5)$ & $n(0.0)$ & Mean score & $n(\textrm{total})$ & $n(1.0)$ & $n(0.5)$ & $n(0.0)$ & Mean score \bigstrut\\
    \hline
    Planet & 67 & 17 & 39 & 11 & 54\% & 69 & 46 & 20 & 3 & 81\% \bigstrut[t]\\
    Star & 71 & 33 & 27 & 11 & 65\% & 70 & 51 & 19 & 0 & 86\% \\
    Solar system & 12 & 4 & 7 & 1 & 63\% & 18 & 16 & 1 & 1 & 92\% \\
    Stellar system & 58 & 45 & 7 & 6 & 84\% & 52 & 50 & 2 & 0 & 98\% \\
    Galaxy & 68 & 15 & 47 & 6 & 57\% & 71 & 28 & 43 & 0 & 70\% \\
    Universe & 71 & 50 & 19 & 2 & 84\% & 70 & 62 & 7 & 1 & 94\% \bigstrut[b]\\
    \hline
    \hline
    \end{tabular}%
\end{table*}
%---------------------------------------------------------------------------

The mean scores for all objects increased pre- to post-course. The biggest gains were seen on ``planet'' and ``solar system'' ($\sim28$ percentage points, equivalent to a $\sim50\%$ increase), with the smallest gain ($\sim10$ percentage points) on ``universe.'' Indeed, when analyzed on a student-by-student basis, almost every individual student's scores on the explaining task increased pre- to post-course. Taking all $6$ objects into account, the class average increased from $68\%$ to $85\%$. 

\textcolor{black}{Interestingly, students' pre-course response rates and scores for the objects ``solar system,'' ``galaxy'' and ``universe'' were very similar to those obtained from a sample of students at five different US institutions using a different instrument, though one where a nearly-identical explaining task was included for the aforesaid three objects, and where a very similar grading scheme was employed \citep{Bailey:2012b}. Preliminary results from a study of pre-service teachers in Norway (see Section \ref{sec:discuss}) also suggest scores consistent with those obtained here. Given the very different sample demographics, it seems that difficulties associated with defining or describing basic astronomical terms might be widespread, and further investigation is merited.}

When analyzed in conjunction with the ranking task, the most prevalent non-scientific description to emerge from the pre-course IAQ data was the idea that stars are small objects fixed on the night sky, which is in turn associated or fixed to the Earth. A few students, all of whom did not speak English as their first language, also did not appear to be familiar with the term ``galaxy''. 

It is worth noting that the majority of students, both pre- and post-course, interpreted ``solar system'' to mean a  general stellar system, and most of those who gave an explanation for the latter also mentioned that the system might have one or more (exo)planets in orbit around it. Similarly, many students interpreted ``planet'' to mean exoplanet. This could be ascribed to the significant coverage exoplanets have enjoyed in recent years in popular media.

Most students' post-course explanations tended to contain more technical detail than their pre-course ones. For example, many more students mentioned thermonuclear fusion and stellar evolution when explaining stars; similarly, students generally gave more detailed characterizations of galaxies (e.g.\ mention of gravitational binding, supermassive black holes at centers, dark matter components, etc.), (exo)planets, and solar/stellar systems.

The wording of the question apparently constrained some of the students' explanations -- the prompt to provide a simple explanation to their ``friend's 12-year old sister'' seems to have led some students to provide overly-simplistic explanations, thus making it difficult to know the extent of their understanding of the entity in question. Positing an audience such as ``a friend who is interested in astronomy'', or similar, may be more productive in future versions of the IAQ.

%---------------------------------------------------------------------------
\subsubsection{\textcolor{black}{Predictors of course success}}\label{sec:corr}
%---------------------------------------------------------------------------
\textcolor{black}{A key aim of the research was to try to identify as early as possible the students who might have difficulty with the course, so that suitable interventions could be put into place. The most promising individual predictor of overall course success was found to be students' pre-course scores on the explaining task (see Sec.\ \ref{sec:Q8b}, and Q8b in Appendix \ref{sec:IAQ-pre}). The Pearson correlation coefficient between students' pre-course explaining task scores (the mean scores in Table \ref{tab:Q8exp}) and final grades was found to be $r=0.59$, with $p$-value $p\ll0.001$ \footnote{The null hypothesis was $r=0$, with alternative hypothesis $r\neq0$.}, indicating a linear correlation significantly different from zero. Overall, students who did better on the pre-course explaining task generally did better on the same explaining task post-course, and more importantly, also obtained better grades for the course as a whole.}

\textcolor{black}{Students' \emph{post-course} scores on the same task were also found to be correlated strongly with final grades: the Pearson correlation coefficient in this case was $r=0.52$, with $p\ll0.001$. Finally, students' pre- and post-course scores on the same explaining task were also correlated, with Pearson coefficient $r=0.63$, again with $p\ll0.001$.}

\begin{figure*}[t]
\centering
\includegraphics[width=86mm]{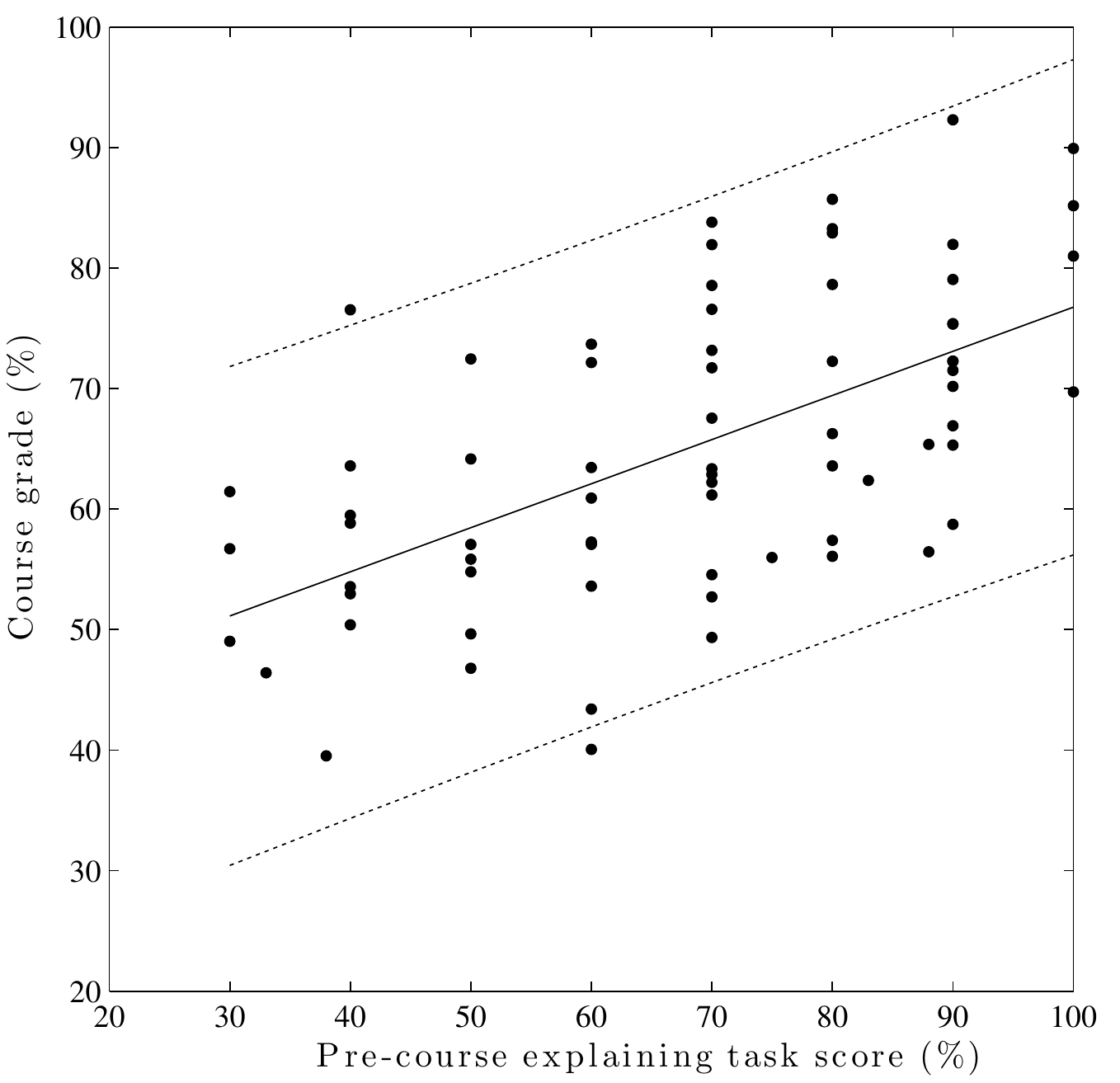}
\includegraphics[width=86mm]{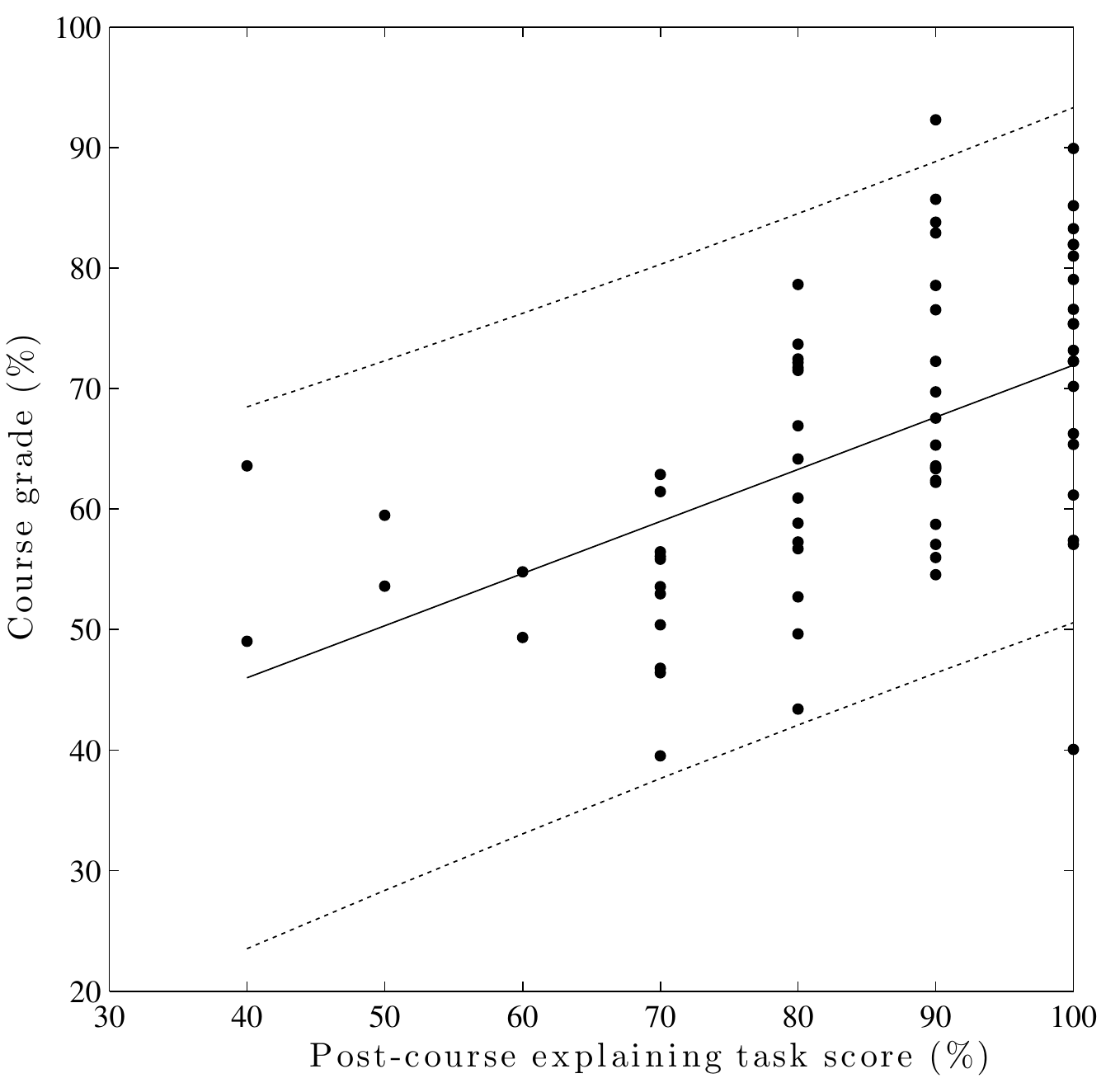}
\caption{Illustration of the linear relationship between final course grades and pre-course scores (left panel) and post-course scores (right panel) on the explaining task. In each plot, the solid line represents the best-fitting linear model for the data, while the dotted lines represent $2\sigma$ prediction bounds for the fit.\label{fig:corr}}
\end{figure*}

\textcolor{black}{The correlations between pre- and post-course scores on the explaining task and final grades are illustrated in Fig.\ \ref{fig:corr}.}

%--------------------------------------------------------------------------------------------------
\section{Discussion and conclusions} \label{sec:discuss}
%--------------------------------------------------------------------------------------------------
The broad aim of this study was to obtain a better understanding of the AST1000F students and their perspectives, in order to improve the course. The instrument we constructed, the IAQ, sought to probe ideas along three broad axes: student motivation and expectations, astronomy content, and worldview. The instrument used a small number of representative questions as a way to sample each of these three areas. Within the limitations of such a brief instrument, a number of potentially-useful insights emerged.

Firstly, contrary to some expectations based on anecdotal evidence, the overwhelming majority of students in the class indicated that an interest in astronomy was their primary reason for doing the course (rather than it being an easy elective); this included many students expressing a long-standing interest in the field, or an interest in a specific topic within the field. The post-course data showed that this interest had been maintained to a high degree, and that students' expectations regarding the content of the course were largely fulfilled.

%-------------------------------------------
% Figure: Correlation stuff
%-------------------------------------------
%-------------------------------------------

The findings based on questions relating to astronomy content indicated that learning gains were made in several conceptual areas, and that students had developed a more nuanced view of the nature of astronomy. However, it also emerged that some concepts (e.g.\ Newtonian gravity) were not well-understood by all students, and that these topics would require further attention in future. 

While only two issues were probed on the worldview axis, it was interesting to note that in the case where a specific intervention was made (addressing astrology), this intervention appeared to change students' thinking. On the other hand, in the case of an issue which was largely addressed implicitly throughout the course, i.e.\ theories and the scientific method, the data from the IAQ were more difficult to interpret unambiguously, although there was an apparent positive change in attitude towards a scientific theory. The broader difficulties of interpretation could be tied to the inherent complexity of the topic, the lack of explicit treatment compared with the previous case, or the way in which the question was changed from pre-course to post-course. Further investigation in this area seems warranted as worldview affects students' approach to science as a whole, and it has become increasingly clear that introductory science courses could play a positive role in addressing broader attitudinal and societal issues without compromising the scientific content of the course \cite{Matthews:2009,Wallace:2013}. Bringing these issues to the fore might also offer opportunities for moving away from binary thinking towards more sophisticated, evidence-based reasoning \cite{Perry:1998}, for which there is little opportunity in the teaching of traditional introductory science courses \cite{Finster:1989,Finster:1991,Kagee:2010}.

%[[Comment re consistency with NBT findings]]

\textcolor{black}{Lastly, we identified a significant correlation between students' scores on the pre-course explaining task and their final grades for the course. Should this correlation hold up for future cohorts, it would suggest using the pre-course results for early intervention on a student-by-student basis. However, it is interesting to note that the explaining task is essentially a writing task, and given that language is often a broad indicator of educational disadvantage, it might be more appropriate in some cases to encourage students to follow the extended-degree route.}

\textcolor{black}{The IAQ has thus proved} to be a rich source of data providing insight into many issues pertaining to students in our first-year astronomy course. Despite the small number of questions, we were able to probe three broad areas relating to student engagement with a scientific course. While analyzing free-response writing is a non-trivial exercise, it is clear to us at this point that such a wealth of information is not easily obtained with a multiple-choice format alone. It is also interesting to note that in many cases, students chose particular options which were not necessarily consistent with their more detailed expositions. Nevertheless, it would be useful to investigate the possibility of generating an MCQ-instrument, as this would allow for more detailed probing along the three axes in question, and would make it feasible to administer the instrument on a large scale. \textcolor{black}{An important preliminary step would be to conduct interviews with a future cohort of students to probe the extent to which the questions are understood as intended.}

The present work is being extended through two further studies, with distinctly different samples. One study involves candidates for the NASSP postgraduate bridging programme; the second involves pre-service teachers based in Norway, using an updated version of the IAQ translated into Norwegian.

%--------------------------------------------------------------------------------------------------
\begin{acknowledgments}
%--------------------------------------------------------------------------------------------------
The authors acknowledge the Academic Development Programme at the University of Cape Town for financial support, and wish to thank Mpeli Takane for assisting with aspects of the analysis. S.A.\ would also like to thank Ed Prather, Benjamin Mendelsohn, and Colin Wallace for a series of inspiring discussions.
%--------------------------------------------------------------------------------------------------
\end{acknowledgments}
%--------------------------------------------------------------------------------------------------
%--------------------------------------------------------------------------------------------------
\appendix

%--------------------------------------------------------------------------------------------------
\section{The pre-course IAQ}\label{sec:IAQ-pre}
%--------------------------------------------------------------------------------------------------

Each question (Q1, Q2, and so on) appeared on a separate page.

\begin{description}
	\item[Q1a] Why did you decide to do the introductory astronomy (AST1000F) course? [10 lines provided for answer.]
	
	\item[Q1b] How difficult do you expect the AST1000F course is going to be? Circle a number from 1 to 5, where 1 means ``extremely easy'' and 5 means ``extremely difficult''. [Boxes with numbers 1 to 5 provided.]
	
	\item[Q2] When the word ``sport'' is mentioned, the following might come to mind: \emph{soccer, tennis, swimming, exercise, keeping fit, injuries, fun, Olympics, Bafana Bafana,} etc. 
	
	Write down all the words or phrases that come to mind when you hear the word ``astronomy''. [Large unruled box filling most of an A4~page provided.]
	
	\item[Q3] A group of students is having an argument about \emph{astronomy} and \emph{astrology}. Student A: ``There is no difference between astronomy and astrology. They are both about the stars.'' Student B: ``Actually, there \emph{is} a difference between astronomy and astrology, but both are .'' Student C: ``I don't agree with either of you!''
	
	With whom do you \emph{most closely} agree: student A, B, or C? (Circle only one letter.) [Boxes with letters A, B, and C provided.]
	
	Explain \emph{in detail} why you chose this option.	[10 lines provided for answer.]

	\item[Q4] A group of students is having an argument about the universe. Student A: ``The universe definitely started with the Big Bang.'' Student B: ``That is not true! The Big Bang is nothing more than a theory.'' Student C: ``I don't agree with either of you!''
	
	With whom do you \emph{most closely} agree: student A, B, or C? (Circle only one letter.) [Boxes with letters A, B, and C provided.]
	
	Explain \emph{in detail} why you chose this option.	[10 lines provided for answer.]

	\item[Q5] A group of students is having an argument about gravity in space. Student A: ``If you drop a pen on the moon it will float forever'' Student B: ``No! You need to be much further away from Earth for gravity to disappear.'' Student C: ``I don't agree with either of you!''
	
	With whom do you \emph{most closely} agree: student A, B, or C? (Circle only one letter.) [Boxes with letters A, B, and C provided.]
	
	Explain \emph{in detail} why you chose this option.	[10 lines provided for answer.]

	\item[Q6] A group of students is having an argument about \emph{radiation}. Student A: ``Radiation is a form of electromagnetic waves.'' Student B: ``No! Radiation consists of nuclear particles.'' Student C: ``I don't agree with either of you!''
	
	With whom do you \emph{most closely} agree: student A, B, or C? (Circle only one letter.) [Boxes with letters A, B, and C provided.]
	
	Explain \emph{in detail} why you chose this option.	[10 lines provided for answer.]
	
		\item[Q7] A group of students is arguing about the Square Kilometre Array (SKA) \emph{radio telescope} that is being built in the Karoo. Student A: ``The SKA will collect \emph{sound waves} from outer space.'' Student B: ``No! It will actually detect \emph{light}!'' Student C: ``I don't agree with either of you!''
	
	With whom do you \emph{most closely} agree: student A, B, or C? (Circle only one letter.) [Boxes with letters A, B, and C provided.]
	
	Explain \emph{in detail} why you chose this option.	[10 lines provided for answer.]
	
	\item[Q8a] Rank the following from smallest to largest: \emph{galaxy; planet; star; universe; solar system}. [5 numbered lines provided, with ``smallest'' in parenthesis next to number 1, and ``largest'' in parenthesis next to number 5.]
	
	\item[Q8b] Now explain to your friend's 12 year old sister what is meant by each of the these. [2 lines provided for each item.]
\end{description}

%--------------------------------------------------------------------------------------------------
\section{The post-course IAQ}\label{sec:IAQ-post}
%--------------------------------------------------------------------------------------------------

Questions 3--6 appeared two to a page; other questions appeared each on a separate page.

\begin{description}
	\item[Q1a] To what extent did the topics covered in this course (AST1000F) meet the expectations you had at the start of the course? Circle a number from 1 to 5, where 1 means ``not at all what I expected'' and 5 means ``exactly what I expected.'' [Boxes with numbers 1 to 5 provided.]
	
	\item[Q1b] How interesting did you find this course? Circle a number from 1 to 5, where 1 means ``not at all interesting'' and 5 means ``extremely interesting.'' [Boxes with numbers 1 to 5 provided.]
	
	\item[Q2] Identical to Q2 in pre-course IAQ.
	
	\item[Q3] A group of students is having an argument about \emph{astronomy} and \emph{astrology}. Student A: ``Astronomy is a useful scientific discipline, whereas there is no evidence to support any of the claims made by astrology, which is mere superstition.'' Student B: ``Nonsense! There is plenty of evidence to support astrology, and it can be just as useful as astronomy.'' Student C: ``Ultimately, astronomy and astrology are both belief systems. Either one of them can be useful, if you choose to believe in it.''
	
	With whom do you \emph{most closely} agree: student A, B, or C? (Circle only one letter.) [Boxes with letters A, B, and C provided.]
	
	\item[Q4] A group of students is having an argument about the origin of the universe. Student A: ``The Big Bang is nothing more than a theory. There is no real evidence that it happened!'' Student B: ``I think you've misunderstood the word ‘theory'. There is strong evidence that the universe started with the Big Bang.'' Student C: ``While there may be evidence for the Big Bang, I am not convinced that is how the universe started.'' Student D: ``Whether or not the Big Bang happened, I believe that God created the universe.''
	
	With whom do you \emph{most closely} agree: student A, B, C, or D? (Circle only one letter.) [Boxes with letters A, B, C, and D provided.]
	
	\item[Q5] A group of students is having the following argument. Student A: ``If you drop a pen on the moon, it will fall to the surface of the moon.'' Student B: ``No! It will not fall down, it will simply float.'' Student C: ``Actually, the pen will drift away from the moon and fall towards the Earth.''
	
	With whom do you \emph{most closely} agree: student A, B, or C? (Circle only one letter.) [Boxes with letters A, B, and C provided.]
	
	\item[Q6] Your friend says to you: ``I heard the astronomy lecturer use the term `radiation' in class, but I wasn't sure what she meant.'' Write down what you will say to your friend in order to enlighten her. [Large unruled box filling most of an A4~page provided.]

	\item[Q7] A group of students is discussing the Square Kilometre Array (SKA) \emph{radio telescope}. Student A: ``Like all telescopes, it will collect and amplify light that is too faint for our eyes to see.'' Student B: ``No! Radio telescopes collect sound waves from outer space.'' Student C: ``I think radio telescopes collect a particular type of electromagnetic wave.''
	
	With whom do you \emph{most closely} agree: student A, B, or C? (Circle only one letter.) [Boxes with letters A, B, and C provided.]

	\item[Q8a] Identical to Q8a in pre-course IAQ.
	
	\item[Q8b] Identical to Q8b in pre-course IAQ.
\end{description}

%--------------------------------------------------------------------------------------------------
\section{Full results for the explaining task}\label{sec:exp-full}
%--------------------------------------------------------------------------------------------------

Table \ref{tab:Q8exp-full} presents the results for the explaining task for the \emph{full} pre- and post-course samples (Sec.\ \ref{sec:Q8b} presents results only for the smaller sample of matched students who answered the question both pre- and post-course). For each object, the total number of students answering the question, the number of students scoring $0.0$, $0.5$ or $1.0$ out of $1.0$, and the mean score for the object, is shown; blank responses were not assigned scores. Note that many students interpreted ``solar system'' as ``stellar system''; scores for the two different possible interpretations are shown separately.

As with the matched sample, the mean scores for all objects increased pre- to post-course, with the biggest gains once again being seen for ``planet'' and ``solar system''. Taking all $6$ objects into account, the class average rose from $65\%$ ($n=79$) to $84\%$ ($n=91$) pre-course to post-course, which is consistent with the $68\%$ to $85\%$ increase seen for the $n=71$ sample of matched students.
%---------------------------------------------------------------------------
\begin{table*}[tbh!]
	\caption{Student scores on the explaining task, for the full pre- and post-course samples.}
	\label{tab:Q8exp-full}
    \begin{tabular}{ccccccccccc}
    \hline
    \hline
      & \multicolumn{5}{c}{Pre-course ($n=79$)} & \multicolumn{5}{c}{Post-course ($n=91$)} \bigstrut\\
    \hline
    Object & $n(\textrm{total})$ & $n(1.0)$ & $n(0.5)$ & $n(0.0)$ & Mean score & $n(\textrm{total})$ & $n(1.0)$ & $n(0.5)$ & $n(0.0)$ & Mean score \bigstrut\\
    \hline
    Planet & 75 & 17 & 43 & 15 & 54\% & 87 & 57 & 25 & 5 & 80\% \bigstrut[t]\\
    Star & 79 & 36 & 30 & 13 & 65\% & 88 & 62 & 25 & 1 & 85\% \\
    Solar system & 17 & 4 & 8 & 5 & 47\% & 24 & 19 & 1 & 4 & 81\% \\
    Stellar system & 61 & 47 & 8 & 6 & 84\% & 66 & 61 & 2 & 3 & 94\% \\
    Galaxy & 75 & 15 & 51 & 9 & 54\% & 88 & 38 & 50 & 0 & 72\% \\
    Universe & 79 & 52 & 23 & 4 & 80\% & 87 & 74 & 11 & 2 & 91\% \bigstrut[b]\\
    \hline
    \hline
    \end{tabular}%
\end{table*}
%---------------------------------------------------------------------------

%--------------------------------------------------------------------------------------------------
\bibliography{IAQ-paper}

\begin{thebibliography}{81}%
\makeatletter
\providecommand \@ifxundefined [1]{%
 \@ifx{#1\undefined}
}%
\providecommand \@ifnum [1]{%
 \ifnum #1\expandafter \@firstoftwo
 \else \expandafter \@secondoftwo
 \fi
}%
\providecommand \@ifx [1]{%
 \ifx #1\expandafter \@firstoftwo
 \else \expandafter \@secondoftwo
 \fi
}%
\providecommand \natexlab [1]{#1}%
\providecommand \enquote  [1]{``#1''}%
\providecommand \bibnamefont  [1]{#1}%
\providecommand \bibfnamefont [1]{#1}%
\providecommand \citenamefont [1]{#1}%
\providecommand \href@noop [0]{\@secondoftwo}%
\providecommand \href [0]{\begingroup \@sanitize@url \@href}%
\providecommand \@href[1]{\@@startlink{#1}\@@href}%
\providecommand \@@href[1]{\endgroup#1\@@endlink}%
\providecommand \@sanitize@url [0]{\catcode `\\12\catcode `\$12\catcode
  `\&12\catcode `\#12\catcode `\^12\catcode `\_12\catcode `\%12\relax}%
\providecommand \@@startlink[1]{}%
\providecommand \@@endlink[0]{}%
\providecommand \url  [0]{\begingroup\@sanitize@url \@url }%
\providecommand \@url [1]{\endgroup\@href {#1}{\urlprefix }}%
\providecommand \urlprefix  [0]{URL }%
\providecommand \Eprint [0]{\href }%
\providecommand \doibase [0]{http://dx.doi.org/}%
\providecommand \selectlanguage [0]{\@gobble}%
\providecommand \bibinfo  [0]{\@secondoftwo}%
\providecommand \bibfield  [0]{\@secondoftwo}%
\providecommand \translation [1]{[#1]}%
\providecommand \BibitemOpen [0]{}%
\providecommand \bibitemStop [0]{}%
\providecommand \bibitemNoStop [0]{.\EOS\space}%
\providecommand \EOS [0]{\spacefactor3000\relax}%
\providecommand \BibitemShut  [1]{\csname bibitem#1\endcsname}%
\let\auto@bib@innerbib\@empty
%</preamble>
\bibitem [{\citenamefont {{The South African Department of Arts, Culture,
  Science and Technology}}(1996)}]{whitepaper:1996}%
  \BibitemOpen
  \bibfield  {author} {\bibinfo {author} {\bibnamefont {{The South African
  Department of Arts, Culture, Science and Technology}}},\ }\href
  {http://www.esastap.org.za/download/st_whitepaper_sep1996.pdf} {\enquote
  {\bibinfo {title} {{White Paper on Science \& Technology}},}\ } (\bibinfo
  {year} {1996})\BibitemShut {NoStop}%
\bibitem [{\citenamefont {{The Government of the Republic of South
  Africa}}(2002)}]{rdstrat:2002}%
  \BibitemOpen
  \bibfield  {author} {\bibinfo {author} {\bibnamefont {{The Government of the
  Republic of South Africa}}},\ }\href
  {www.info.gov.za/otherdocs/2002/rd_strat.pdf} {\enquote {\bibinfo {title}
  {{South Africa's National Research and Development Strategy}},}\ } (\bibinfo
  {year} {2002})\BibitemShut {NoStop}%
\bibitem [{\citenamefont {{Whitelock}}(2008{\natexlab{a}})}]{Whitelock:2008a}%
  \BibitemOpen
  \bibfield  {author} {\bibinfo {author} {\bibfnamefont {P.}~\bibnamefont
  {{Whitelock}}},\ }in\ \href {\doibase 10.1063/1.2905138} {\emph {\bibinfo
  {booktitle} {Proceedings of the National Society Of Black Physicists}}},\
  \bibinfo {series} {American Institute of Physics Conference Series}, Vol.\
  \bibinfo {volume} {991},\ \bibinfo {editor} {edited by\ \bibinfo {editor}
  {\bibfnamefont {H.~M.}\ \bibnamefont {{Oluseyi}}}}\ (\bibinfo {year} {2008})\
  pp.\ \bibinfo {pages} {37--46},\ \Eprint {http://arxiv.org/abs/0707.0921}
  {arXiv:0707.0921} \BibitemShut {NoStop}%
\bibitem [{Note1()}]{Note1}%
  \BibitemOpen
  \bibinfo {note} {In South African parlance, the term `Faculty of Science'
  refers to an organizational structure and not the staff within the structure.
  The analogous American terminology would be `School of Science' or `College
  of Science.' `Staff' is usually used in a general sense and includes
  `faculty' as used in the American sense.}\BibitemShut {Stop}%
\bibitem [{Note2()}]{Note2}%
  \BibitemOpen
  \bibinfo {note} {Compared to bachelors' degrees in the USA, South African
  degrees tend to be much more rigid. A Bachelor of Science degree in South
  Africa nominally comprises the equivalent of $9$ full-year courses spread
  over three years: typically, four first-year courses, three second-year
  courses and two third-year (major) courses. For example, first-year courses
  for a student majoring in Astrophysics and Physics could be Physics I,
  Mathematics I, Applied Mathematics I, Astronomy I (half-course) and
  Statistics I (half-course). Second-year courses would consist of Astrophysics
  II, Physics II, and Mathematics II or Applied Mathematics II, while
  third-year courses would be Astrophysics III and Physics III. The three-year
  Bachelor of Science degree is followed by a one-year Bachelor of Science
  (Honours) degree, comparable to the first year of a PhD programme in the USA.
  Typically, further postgraduate study will entail a research Master of
  Science degree (nominally two years), followed by a PhD (nominally three
  years).}\BibitemShut {Stop}%
\bibitem [{\citenamefont {{Whitelock}}(2008{\natexlab{b}})}]{Whitelock:2008b}%
  \BibitemOpen
  \bibfield  {author} {\bibinfo {author} {\bibfnamefont {P.~A.}\ \bibnamefont
  {{Whitelock}}},\ }in\ \href@noop {} {\emph {\bibinfo {booktitle} {First
  Middle East-Africa Regional IAU Meeting}}}\ (\bibinfo {year}
  {2008})\BibitemShut {NoStop}%
\bibitem [{Note3()}]{Note3}%
  \BibitemOpen
  \bibinfo {note} {The terms `black' and `African' are often used
  interchangeably in the South African context. In this paper, the term `black'
  is used to include all previously disenfranchised groups, in particular those
  previously categorized in South Africa as African, Colored, and Indian (of
  Asian origin).}\BibitemShut {Stop}%
\bibitem [{\citenamefont {{Partridge}}\ and\ \citenamefont
  {{Greenstein}}(2003)}]{Partridge:2003}%
  \BibitemOpen
  \bibfield  {author} {\bibinfo {author} {\bibfnamefont {B.}~\bibnamefont
  {{Partridge}}}\ and\ \bibinfo {author} {\bibfnamefont {G.}~\bibnamefont
  {{Greenstein}}},\ }\href@noop {} {\bibfield  {journal} {\bibinfo  {journal}
  {Astronomy Education Review}\ }\textbf {\bibinfo {volume} {2}},\ \bibinfo
  {pages} {46} (\bibinfo {year} {2003})}\BibitemShut {NoStop}%
\bibitem [{\citenamefont {{Zeilik}}\ and\ \citenamefont
  {{Morris-Dueer}}(2004)}]{Zeilik:2004}%
  \BibitemOpen
  \bibfield  {author} {\bibinfo {author} {\bibfnamefont {M.}~\bibnamefont
  {{Zeilik}}}\ and\ \bibinfo {author} {\bibfnamefont {V.~J.}\ \bibnamefont
  {{Morris-Dueer}}},\ }\href@noop {} {\bibfield  {journal} {\bibinfo  {journal}
  {Astronomy Education Review}\ }\textbf {\bibinfo {volume} {3}},\ \bibinfo
  {pages} {61} (\bibinfo {year} {2004})}\BibitemShut {NoStop}%
\bibitem [{\citenamefont {{Chaisson}}\ and\ \citenamefont
  {{McMillan}}(2010)}]{TEXTBOOK:2010}%
  \BibitemOpen
  \bibfield  {author} {\bibinfo {author} {\bibfnamefont {E.}~\bibnamefont
  {{Chaisson}}}\ and\ \bibinfo {author} {\bibfnamefont {S.}~\bibnamefont
  {{McMillan}}},\ }\href@noop {} {\emph {\bibinfo {title} {{Astronomy
  Today}}}},\ \bibinfo {edition} {7th}\ ed.\ (\bibinfo  {publisher} {Pearson},\
  \bibinfo {year} {2010})\BibitemShut {NoStop}%
\bibitem [{\citenamefont {Prather}\ \emph {et~al.}(2008)\citenamefont
  {Prather}, \citenamefont {Slater}, \citenamefont {Adams},\ and\ \citenamefont
  {Brissenden}}]{Prather:2008}%
  \BibitemOpen
  \bibfield  {author} {\bibinfo {author} {\bibfnamefont {E.~E.}\ \bibnamefont
  {Prather}}, \bibinfo {author} {\bibfnamefont {T.~F.}\ \bibnamefont {Slater}},
  \bibinfo {author} {\bibfnamefont {J.~P.}\ \bibnamefont {Adams}}, \ and\
  \bibinfo {author} {\bibfnamefont {G.}~\bibnamefont {Brissenden}},\
  }\href@noop {} {\emph {\bibinfo {title} {Lecture-tutorials for introductory
  astronomy}}},\ \bibinfo {edition} {2nd}\ ed.\ (\bibinfo  {publisher}
  {Addison-Wesley},\ \bibinfo {address} {San Francisco, CA},\ \bibinfo {year}
  {2008})\ \bibinfo {note} {addison-Wesley Series in Educational
  Innovation}\BibitemShut {NoStop}%
\bibitem [{\citenamefont {{Marschall}}\ \emph {et~al.}(2000)\citenamefont
  {{Marschall}}, \citenamefont {{Snyder}},\ and\ \citenamefont
  {{Cooper}}}]{CLEA:2000}%
  \BibitemOpen
  \bibfield  {author} {\bibinfo {author} {\bibfnamefont {L.~A.}\ \bibnamefont
  {{Marschall}}}, \bibinfo {author} {\bibfnamefont {G.~A.}\ \bibnamefont
  {{Snyder}}}, \ and\ \bibinfo {author} {\bibfnamefont {P.~R.}\ \bibnamefont
  {{Cooper}}},\ }\href {\doibase 10.1119/1.1341943} {\bibfield  {journal}
  {\bibinfo  {journal} {The Physics Teacher}\ }\textbf {\bibinfo {volume}
  {38}},\ \bibinfo {pages} {536} (\bibinfo {year} {2000})}\BibitemShut
  {NoStop}%
\bibitem [{\citenamefont {{Lintott}}\ \emph {et~al.}(2011)\citenamefont
  {{Lintott}}, \citenamefont {{Schwamb}}, \citenamefont {{Fischer}},
  \citenamefont {{Giguere}}, \citenamefont {{Lynn}}, \citenamefont {{Brewer}},
  \citenamefont {{Schawinski}}, \citenamefont {{Simpson}}, \citenamefont
  {{Smith}},\ and\ \citenamefont {{Spronck}}}]{HUNTERS:2011}%
  \BibitemOpen
  \bibfield  {author} {\bibinfo {author} {\bibfnamefont {C.~J.}\ \bibnamefont
  {{Lintott}}}, \bibinfo {author} {\bibfnamefont {M.~E.}\ \bibnamefont
  {{Schwamb}}}, \bibinfo {author} {\bibfnamefont {D.~A.}\ \bibnamefont
  {{Fischer}}}, \bibinfo {author} {\bibfnamefont {M.~J.}\ \bibnamefont
  {{Giguere}}}, \bibinfo {author} {\bibfnamefont {S.}~\bibnamefont {{Lynn}}},
  \bibinfo {author} {\bibfnamefont {J.~M.}\ \bibnamefont {{Brewer}}}, \bibinfo
  {author} {\bibfnamefont {K.}~\bibnamefont {{Schawinski}}}, \bibinfo {author}
  {\bibfnamefont {R.~J.}\ \bibnamefont {{Simpson}}}, \bibinfo {author}
  {\bibfnamefont {A.}~\bibnamefont {{Smith}}}, \ and\ \bibinfo {author}
  {\bibfnamefont {J.}~\bibnamefont {{Spronck}}},\ }in\ \href@noop {} {\emph
  {\bibinfo {booktitle} {EPSC-DPS Joint Meeting 2011}}}\ (\bibinfo {year}
  {2011})\ p.\ \bibinfo {pages} {1226}\BibitemShut {NoStop}%
\bibitem [{\citenamefont {{Raddick}}\ \emph {et~al.}(2007)\citenamefont
  {{Raddick}}, \citenamefont {{Lintott}}, \citenamefont {{Schawinski}},
  \citenamefont {{Thomas}}, \citenamefont {{Nichol}}, \citenamefont
  {{Andreescu}}, \citenamefont {{Bamford}}, \citenamefont {{Land}},
  \citenamefont {{Murray}}, \citenamefont {{Slosar}}, \citenamefont {{Szalay}},
  \citenamefont {{Vandenberg}},\ and\ \citenamefont {{Galaxy Zoo
  Team}}}]{ZOO:2007}%
  \BibitemOpen
  \bibfield  {author} {\bibinfo {author} {\bibfnamefont {J.}~\bibnamefont
  {{Raddick}}}, \bibinfo {author} {\bibfnamefont {C.~J.}\ \bibnamefont
  {{Lintott}}}, \bibinfo {author} {\bibfnamefont {K.}~\bibnamefont
  {{Schawinski}}}, \bibinfo {author} {\bibfnamefont {D.}~\bibnamefont
  {{Thomas}}}, \bibinfo {author} {\bibfnamefont {R.~C.}\ \bibnamefont
  {{Nichol}}}, \bibinfo {author} {\bibfnamefont {D.}~\bibnamefont
  {{Andreescu}}}, \bibinfo {author} {\bibfnamefont {S.}~\bibnamefont
  {{Bamford}}}, \bibinfo {author} {\bibfnamefont {K.~R.}\ \bibnamefont
  {{Land}}}, \bibinfo {author} {\bibfnamefont {P.}~\bibnamefont {{Murray}}},
  \bibinfo {author} {\bibfnamefont {A.}~\bibnamefont {{Slosar}}}, \bibinfo
  {author} {\bibfnamefont {A.~S.}\ \bibnamefont {{Szalay}}}, \bibinfo {author}
  {\bibfnamefont {J.}~\bibnamefont {{Vandenberg}}}, \ and\ \bibinfo {author}
  {\bibnamefont {{Galaxy Zoo Team}}},\ }in\ \href@noop {} {\emph {\bibinfo
  {booktitle} {American Astronomical Society Meeting Abstracts}}},\ \bibinfo
  {series} {Bulletin of the American Astronomical Society}, Vol.~\bibinfo
  {volume} {39}\ (\bibinfo {year} {2007})\ p.\ \bibinfo {pages}
  {892}\BibitemShut {NoStop}%
\bibitem [{Note4()}]{Note4}%
  \BibitemOpen
  \bibinfo {note} {Grading of all inputs is done on a numerical rather than
  alphabetical scale, with $50\%$ constituting a passing grade.}\BibitemShut
  {Stop}%
\bibitem [{\citenamefont {{Fraknoi}}(2002)}]{Fraknoi:2002}%
  \BibitemOpen
  \bibfield  {author} {\bibinfo {author} {\bibfnamefont {A.}~\bibnamefont
  {{Fraknoi}}},\ }\href@noop {} {\bibfield  {journal} {\bibinfo  {journal}
  {Astronomy Education Review}\ }\textbf {\bibinfo {volume} {1}},\ \bibinfo
  {pages} {121} (\bibinfo {year} {2002})}\BibitemShut {NoStop}%
\bibitem [{\citenamefont {Harber}(2001)}]{Harber:2001}%
  \BibitemOpen
  \bibfield  {author} {\bibinfo {author} {\bibfnamefont {C.}~\bibnamefont
  {Harber}},\ }\href@noop {} {\emph {\bibinfo {title} {State of Transition:
  Post-Apartheid Educational Reform in South Africa}}},\ Monographs in
  International Education\ (\bibinfo  {publisher} {Symposium Books},\ \bibinfo
  {year} {2001})\BibitemShut {NoStop}%
\bibitem [{\citenamefont {Brock-Utne}(2005)}]{Utne:2005}%
  \BibitemOpen
  \bibfield  {author} {\bibinfo {author} {\bibfnamefont {B.}~\bibnamefont
  {Brock-Utne}},\ }in\ \href {\doibase 10.1007/1-4020-2960-8_34} {\emph
  {\bibinfo {booktitle} {International Handbook on Globalisation, Education and
  Policy Research}}},\ \bibinfo {editor} {edited by\ \bibinfo {editor}
  {\bibfnamefont {J.}~\bibnamefont {Zajda}}, \bibinfo {editor} {\bibfnamefont
  {K.}~\bibnamefont {Freeman}}, \bibinfo {editor} {\bibfnamefont
  {M.}~\bibnamefont {Geo-Jaja}}, \bibinfo {editor} {\bibfnamefont
  {S.}~\bibnamefont {Majhanovic}}, \bibinfo {editor} {\bibfnamefont
  {V.}~\bibnamefont {Rust}}, \bibinfo {editor} {\bibfnamefont {J.}~\bibnamefont
  {Zajda}}, \ and\ \bibinfo {editor} {\bibfnamefont {R.}~\bibnamefont
  {Zajda}}}\ (\bibinfo  {publisher} {Springer},\ \bibinfo {year} {2005})\ pp.\
  \bibinfo {pages} {549--565}\BibitemShut {NoStop}%
\bibitem [{\citenamefont {{The Department of Education of South
  Africa}}(2002)}]{NCSjunior:2002}%
  \BibitemOpen
  \bibfield  {author} {\bibinfo {author} {\bibnamefont {{The Department of
  Education of South Africa}}},\ }\href
  {http://www.education.gov.za/LinkClick.aspx?fileticket=DcpfuTKf6rE=}
  {\enquote {\bibinfo {title} {{Revised National Curriculum Statement for
  Grades R-9 (Schools), Natural Sciences}},}\ } (\bibinfo {year} {2002}),\
  \bibinfo {note} {{ISBN 1-919917-48-9}}\BibitemShut {NoStop}%
\bibitem [{\citenamefont {{The Department of Education of South
  Africa}}(2003)}]{NCSsenior:2003}%
  \BibitemOpen
  \bibfield  {author} {\bibinfo {author} {\bibnamefont {{The Department of
  Education of South Africa}}},\ }\href
  {http://www.education.gov.za/LinkClick.aspx?fileticket=VoSn0yzfkcE=}
  {\enquote {\bibinfo {title} {{National Curriculum Statement, Grades 10-12
  (General), Physical Sciences}},}\ } (\bibinfo {year} {2003}),\ \bibinfo
  {note} {{ISBN 1-919975-61-6}}\BibitemShut {NoStop}%
\bibitem [{\citenamefont {{Allie}}\ \emph {et~al.}(1998)\citenamefont
  {{Allie}}, \citenamefont {{Buffler}}, \citenamefont {{Campbell}},\ and\
  \citenamefont {{Lubben}}}]{Allie:1998}%
  \BibitemOpen
  \bibfield  {author} {\bibinfo {author} {\bibfnamefont {S.}~\bibnamefont
  {{Allie}}}, \bibinfo {author} {\bibfnamefont {A.}~\bibnamefont {{Buffler}}},
  \bibinfo {author} {\bibfnamefont {B.}~\bibnamefont {{Campbell}}}, \ and\
  \bibinfo {author} {\bibfnamefont {F.}~\bibnamefont {{Lubben}}},\ }\href
  {\doibase 10.1080/0950069980200405} {\bibfield  {journal} {\bibinfo
  {journal} {International Journal of Science Education}\ }\textbf {\bibinfo
  {volume} {20}},\ \bibinfo {pages} {447} (\bibinfo {year} {1998})}\BibitemShut
  {NoStop}%
\bibitem [{\citenamefont {{Allie}}\ and\ \citenamefont
  {{Buffler}}(1998)}]{Allie:1998b}%
  \BibitemOpen
  \bibfield  {author} {\bibinfo {author} {\bibfnamefont {S.}~\bibnamefont
  {{Allie}}}\ and\ \bibinfo {author} {\bibfnamefont {A.}~\bibnamefont
  {{Buffler}}},\ }\href {\doibase 10.1119/1.18915} {\bibfield  {journal}
  {\bibinfo  {journal} {American Journal of Physics}\ }\textbf {\bibinfo
  {volume} {66}},\ \bibinfo {pages} {613} (\bibinfo {year} {1998})}\BibitemShut
  {NoStop}%
\bibitem [{\citenamefont {Campbell}\ \emph {et~al.}(2000)\citenamefont
  {Campbell}, \citenamefont {Kaunda}, \citenamefont {Allie}, \citenamefont
  {Buffler},\ and\ \citenamefont {Lubben}}]{Campbell:2000}%
  \BibitemOpen
  \bibfield  {author} {\bibinfo {author} {\bibfnamefont {B.}~\bibnamefont
  {Campbell}}, \bibinfo {author} {\bibfnamefont {L.}~\bibnamefont {Kaunda}},
  \bibinfo {author} {\bibfnamefont {S.}~\bibnamefont {Allie}}, \bibinfo
  {author} {\bibfnamefont {A.}~\bibnamefont {Buffler}}, \ and\ \bibinfo
  {author} {\bibfnamefont {F.}~\bibnamefont {Lubben}},\ }\href {\doibase
  10.1002/1098-2736(200010)37:8<839::AID-TEA5>3.0.CO;2-W} {\bibfield  {journal}
  {\bibinfo  {journal} {Journal of Research in Science Teaching}\ }\textbf
  {\bibinfo {volume} {37}},\ \bibinfo {pages} {839} (\bibinfo {year}
  {2000})}\BibitemShut {NoStop}%
\bibitem [{\citenamefont {{Allie}}\ \emph {et~al.}(2003)\citenamefont
  {{Allie}}, \citenamefont {{Buffler}}, \citenamefont {{Campbell}},
  \citenamefont {{Lubben}}, \citenamefont {{Evangelinos}}, \citenamefont
  {{Psillos}},\ and\ \citenamefont {{Valassiades}}}]{Allie:2003}%
  \BibitemOpen
  \bibfield  {author} {\bibinfo {author} {\bibfnamefont {S.}~\bibnamefont
  {{Allie}}}, \bibinfo {author} {\bibfnamefont {A.}~\bibnamefont {{Buffler}}},
  \bibinfo {author} {\bibfnamefont {B.}~\bibnamefont {{Campbell}}}, \bibinfo
  {author} {\bibfnamefont {F.}~\bibnamefont {{Lubben}}}, \bibinfo {author}
  {\bibfnamefont {D.}~\bibnamefont {{Evangelinos}}}, \bibinfo {author}
  {\bibfnamefont {D.}~\bibnamefont {{Psillos}}}, \ and\ \bibinfo {author}
  {\bibfnamefont {O.}~\bibnamefont {{Valassiades}}},\ }\href {\doibase
  10.1119/1.1616479} {\bibfield  {journal} {\bibinfo  {journal} {The Physics
  Teacher}\ }\textbf {\bibinfo {volume} {41}},\ \bibinfo {pages} {394}
  (\bibinfo {year} {2003})}\BibitemShut {NoStop}%
\bibitem [{\citenamefont {{Buffler}}\ \emph {et~al.}(2008)\citenamefont
  {{Buffler}}, \citenamefont {{Allie}},\ and\ \citenamefont
  {{Lubben}}}]{Buffler:2008}%
  \BibitemOpen
  \bibfield  {author} {\bibinfo {author} {\bibfnamefont {A.}~\bibnamefont
  {{Buffler}}}, \bibinfo {author} {\bibfnamefont {S.}~\bibnamefont {{Allie}}},
  \ and\ \bibinfo {author} {\bibfnamefont {F.}~\bibnamefont {{Lubben}}},\
  }\href {\doibase 10.1119/1.3023655} {\bibfield  {journal} {\bibinfo
  {journal} {The Physics Teacher}\ }\textbf {\bibinfo {volume} {46}},\ \bibinfo
  {pages} {539} (\bibinfo {year} {2008})}\BibitemShut {NoStop}%
\bibitem [{\citenamefont {Volkwyn}\ \emph {et~al.}(2008)\citenamefont
  {Volkwyn}, \citenamefont {Allie}, \citenamefont {Buffler},\ and\
  \citenamefont {Lubben}}]{Volkwyn:2008}%
  \BibitemOpen
  \bibfield  {author} {\bibinfo {author} {\bibfnamefont {T.~S.}\ \bibnamefont
  {Volkwyn}}, \bibinfo {author} {\bibfnamefont {S.}~\bibnamefont {Allie}},
  \bibinfo {author} {\bibfnamefont {A.}~\bibnamefont {Buffler}}, \ and\
  \bibinfo {author} {\bibfnamefont {F.}~\bibnamefont {Lubben}},\ }\href
  {\doibase 10.1103/PhysRevSTPER.4.010108} {\bibfield  {journal} {\bibinfo
  {journal} {Phys. Rev. ST Phys. Educ. Res.}\ }\textbf {\bibinfo {volume}
  {4}},\ \bibinfo {pages} {010108} (\bibinfo {year} {2008})}\BibitemShut
  {NoStop}%
\bibitem [{\citenamefont {Lubben}\ \emph {et~al.}(2010)\citenamefont {Lubben},
  \citenamefont {Davidowitz}, \citenamefont {Buffler}, \citenamefont {Allie},\
  and\ \citenamefont {Scott}}]{Lubben:2010}%
  \BibitemOpen
  \bibfield  {author} {\bibinfo {author} {\bibfnamefont {F.}~\bibnamefont
  {Lubben}}, \bibinfo {author} {\bibfnamefont {B.}~\bibnamefont {Davidowitz}},
  \bibinfo {author} {\bibfnamefont {A.}~\bibnamefont {Buffler}}, \bibinfo
  {author} {\bibfnamefont {S.}~\bibnamefont {Allie}}, \ and\ \bibinfo {author}
  {\bibfnamefont {I.}~\bibnamefont {Scott}},\ }\href {\doibase
  http://dx.doi.org/10.1016/j.ijedudev.2009.11.009} {\bibfield  {journal}
  {\bibinfo  {journal} {International Journal of Educational Development}\
  }\textbf {\bibinfo {volume} {30}},\ \bibinfo {pages} {351} (\bibinfo {year}
  {2010})}\BibitemShut {NoStop}%
\bibitem [{\citenamefont {{Nwosu}}\ \emph {et~al.}(2011)\citenamefont
  {{Nwosu}}, \citenamefont {{Demaree}}, \citenamefont {{Li}},\ and\
  \citenamefont {{Allie}}}]{Nwosu:2011}%
  \BibitemOpen
  \bibfield  {author} {\bibinfo {author} {\bibfnamefont {V.}~\bibnamefont
  {{Nwosu}}}, \bibinfo {author} {\bibfnamefont {D.}~\bibnamefont {{Demaree}}},
  \bibinfo {author} {\bibfnamefont {S.}~\bibnamefont {{Li}}}, \ and\ \bibinfo
  {author} {\bibfnamefont {S.}~\bibnamefont {{Allie}}},\ }in\ \href@noop {}
  {\emph {\bibinfo {booktitle} {APS Northwest Section Meeting Abstracts}}}\
  (\bibinfo {year} {2011})\ p.\ \bibinfo {pages} {H1006}\BibitemShut {NoStop}%
\bibitem [{\citenamefont {{Nwosu}}\ \emph {et~al.}(2013)\citenamefont
  {{Nwosu}}, \citenamefont {{Allie}}, \citenamefont {{Demaree}},\ and\
  \citenamefont {{Deacon}}}]{Nwosu:2013}%
  \BibitemOpen
  \bibfield  {author} {\bibinfo {author} {\bibfnamefont {V.}~\bibnamefont
  {{Nwosu}}}, \bibinfo {author} {\bibfnamefont {S.}~\bibnamefont {{Allie}}},
  \bibinfo {author} {\bibfnamefont {D.}~\bibnamefont {{Demaree}}}, \ and\
  \bibinfo {author} {\bibfnamefont {A.}~\bibnamefont {{Deacon}}},\ }in\ \href
  {\doibase 10.1063/1.4789711} {\emph {\bibinfo {booktitle} {American Institute
  of Physics Conference Series}}},\ \bibinfo {series} {American Institute of
  Physics Conference Series}, Vol.\ \bibinfo {volume} {1513},\ \bibinfo
  {editor} {edited by\ \bibinfo {editor} {\bibfnamefont {P.~V.}\ \bibnamefont
  {{Engelhardt}}}, \bibinfo {editor} {\bibfnamefont {A.~D.}\ \bibnamefont
  {{Churukian}}}, \ and\ \bibinfo {editor} {\bibfnamefont {N.~S.}\ \bibnamefont
  {{Rebello}}}}\ (\bibinfo {year} {2013})\ pp.\ \bibinfo {pages}
  {298--301}\BibitemShut {NoStop}%
\bibitem [{\citenamefont {{Wallace}}\ \emph {et~al.}(2013)\citenamefont
  {{Wallace}}, \citenamefont {{Prather}},\ and\ \citenamefont
  {{Mendelsohn}}}]{Wallace:2013}%
  \BibitemOpen
  \bibfield  {author} {\bibinfo {author} {\bibfnamefont {C.~S.}\ \bibnamefont
  {{Wallace}}}, \bibinfo {author} {\bibfnamefont {E.~E.}\ \bibnamefont
  {{Prather}}}, \ and\ \bibinfo {author} {\bibfnamefont {B.~M.}\ \bibnamefont
  {{Mendelsohn}}},\ }\href {\doibase 10.3847/AER2012042} {\bibfield  {journal}
  {\bibinfo  {journal} {Astronomy Education Review}\ }\textbf {\bibinfo
  {volume} {12}},\ \bibinfo {pages} {010101} (\bibinfo {year}
  {2013})}\BibitemShut {NoStop}%
\bibitem [{\citenamefont {{Hudgins}}\ \emph {et~al.}(2006)\citenamefont
  {{Hudgins}}, \citenamefont {{Prather}}, \citenamefont {{Grayson}},\ and\
  \citenamefont {{Smits}}}]{2006AEdRv...5a...1H}%
  \BibitemOpen
  \bibfield  {author} {\bibinfo {author} {\bibfnamefont {D.~W.}\ \bibnamefont
  {{Hudgins}}}, \bibinfo {author} {\bibfnamefont {E.~E.}\ \bibnamefont
  {{Prather}}}, \bibinfo {author} {\bibfnamefont {D.~J.}\ \bibnamefont
  {{Grayson}}}, \ and\ \bibinfo {author} {\bibfnamefont {D.~P.}\ \bibnamefont
  {{Smits}}},\ }\href@noop {} {\bibfield  {journal} {\bibinfo  {journal}
  {Astronomy Education Review}\ }\textbf {\bibinfo {volume} {5}},\ \bibinfo
  {pages} {010000} (\bibinfo {year} {2006})}\BibitemShut {NoStop}%
\bibitem [{\citenamefont {{Hufnagel}}\ \emph {et~al.}(2000)\citenamefont
  {{Hufnagel}}, \citenamefont {{Slater}}, \citenamefont {{Deming}},
  \citenamefont {{Adams}}, \citenamefont {{Adrian}}, \citenamefont {{Brick}},\
  and\ \citenamefont {{Zeilik}}}]{Hufnagel:2000}%
  \BibitemOpen
  \bibfield  {author} {\bibinfo {author} {\bibfnamefont {B.}~\bibnamefont
  {{Hufnagel}}}, \bibinfo {author} {\bibfnamefont {T.~F.}\ \bibnamefont
  {{Slater}}}, \bibinfo {author} {\bibfnamefont {G.}~\bibnamefont {{Deming}}},
  \bibinfo {author} {\bibfnamefont {J.~P.}\ \bibnamefont {{Adams}}}, \bibinfo
  {author} {\bibfnamefont {R.~L.}\ \bibnamefont {{Adrian}}}, \bibinfo {author}
  {\bibfnamefont {C.}~\bibnamefont {{Brick}}}, \ and\ \bibinfo {author}
  {\bibfnamefont {M.}~\bibnamefont {{Zeilik}}},\ }\href {\doibase
  10.1071/AS00152} {\bibfield  {journal} {\bibinfo  {journal} {\pasa}\ }\textbf
  {\bibinfo {volume} {17}},\ \bibinfo {pages} {152} (\bibinfo {year}
  {2000})}\BibitemShut {NoStop}%
\bibitem [{\citenamefont {{Zeilik}}(2002)}]{Zeilik:2002}%
  \BibitemOpen
  \bibfield  {author} {\bibinfo {author} {\bibfnamefont {M.}~\bibnamefont
  {{Zeilik}}},\ }\href@noop {} {\bibfield  {journal} {\bibinfo  {journal}
  {Astronomy Education Review}\ }\textbf {\bibinfo {volume} {1}},\ \bibinfo
  {pages} {46} (\bibinfo {year} {2002})}\BibitemShut {NoStop}%
\bibitem [{\citenamefont {{Hufnagel}}(2002)}]{Hufnagel:2002}%
  \BibitemOpen
  \bibfield  {author} {\bibinfo {author} {\bibfnamefont {B.}~\bibnamefont
  {{Hufnagel}}},\ }\href@noop {} {\bibfield  {journal} {\bibinfo  {journal}
  {Astronomy Education Review}\ }\textbf {\bibinfo {volume} {1}},\ \bibinfo
  {pages} {47} (\bibinfo {year} {2002})}\BibitemShut {NoStop}%
\bibitem [{\citenamefont {{Deming}}(2002)}]{Deming:2002}%
  \BibitemOpen
  \bibfield  {author} {\bibinfo {author} {\bibfnamefont {G.~L.}\ \bibnamefont
  {{Deming}}},\ }\href@noop {} {\bibfield  {journal} {\bibinfo  {journal}
  {Astronomy Education Review}\ }\textbf {\bibinfo {volume} {1}},\ \bibinfo
  {pages} {52} (\bibinfo {year} {2002})}\BibitemShut {NoStop}%
\bibitem [{\citenamefont {{Brogt}}\ \emph {et~al.}(2007)\citenamefont
  {{Brogt}}, \citenamefont {{Sabers}}, \citenamefont {{Prather}}, \citenamefont
  {{Deming}}, \citenamefont {{Hufnagel}},\ and\ \citenamefont
  {{Slater}}}]{Brogt:2007}%
  \BibitemOpen
  \bibfield  {author} {\bibinfo {author} {\bibfnamefont {E.}~\bibnamefont
  {{Brogt}}}, \bibinfo {author} {\bibfnamefont {D.}~\bibnamefont {{Sabers}}},
  \bibinfo {author} {\bibfnamefont {E.~E.}\ \bibnamefont {{Prather}}}, \bibinfo
  {author} {\bibfnamefont {G.~L.}\ \bibnamefont {{Deming}}}, \bibinfo {author}
  {\bibfnamefont {B.}~\bibnamefont {{Hufnagel}}}, \ and\ \bibinfo {author}
  {\bibfnamefont {T.~F.}\ \bibnamefont {{Slater}}},\ }\href@noop {} {\bibfield
  {journal} {\bibinfo  {journal} {Astronomy Education Review}\ }\textbf
  {\bibinfo {volume} {6}},\ \bibinfo {pages} {25} (\bibinfo {year}
  {2007})}\BibitemShut {NoStop}%
\bibitem [{\citenamefont {{Bardar}}\ \emph
  {et~al.}(2006{\natexlab{a}})\citenamefont {{Bardar}}, \citenamefont
  {{Prather}}, \citenamefont {{Brecher}},\ and\ \citenamefont
  {{Slater}}}]{Bardar:2006}%
  \BibitemOpen
  \bibfield  {author} {\bibinfo {author} {\bibfnamefont {E.~M.~W.}\
  \bibnamefont {{Bardar}}}, \bibinfo {author} {\bibfnamefont {E.~E.}\
  \bibnamefont {{Prather}}}, \bibinfo {author} {\bibfnamefont {K.}~\bibnamefont
  {{Brecher}}}, \ and\ \bibinfo {author} {\bibfnamefont {T.~F.}\ \bibnamefont
  {{Slater}}},\ }\href@noop {} {\bibfield  {journal} {\bibinfo  {journal}
  {Astronomy Education Review}\ }\textbf {\bibinfo {volume} {4}},\ \bibinfo
  {pages} {20} (\bibinfo {year} {2006}{\natexlab{a}})}\BibitemShut {NoStop}%
\bibitem [{\citenamefont {{Bardar}}\ \emph
  {et~al.}(2006{\natexlab{b}})\citenamefont {{Bardar}}, \citenamefont
  {{Prather}}, \citenamefont {{Brecher}},\ and\ \citenamefont
  {{Slater}}}]{Bardar:2006b}%
  \BibitemOpen
  \bibfield  {author} {\bibinfo {author} {\bibfnamefont {E.~M.}\ \bibnamefont
  {{Bardar}}}, \bibinfo {author} {\bibfnamefont {E.~E.}\ \bibnamefont
  {{Prather}}}, \bibinfo {author} {\bibfnamefont {K.}~\bibnamefont
  {{Brecher}}}, \ and\ \bibinfo {author} {\bibfnamefont {T.~F.}\ \bibnamefont
  {{Slater}}},\ }\href@noop {} {\bibfield  {journal} {\bibinfo  {journal}
  {Astronomy Education Review}\ }\textbf {\bibinfo {volume} {5}},\ \bibinfo
  {pages} {103} (\bibinfo {year} {2006}{\natexlab{b}})}\BibitemShut {NoStop}%
\bibitem [{\citenamefont {{Bailey}}(2007)}]{Bailey:2007}%
  \BibitemOpen
  \bibfield  {author} {\bibinfo {author} {\bibfnamefont {J.~M.}\ \bibnamefont
  {{Bailey}}},\ }\href {\doibase 10.3847/AER2007028} {\bibfield  {journal}
  {\bibinfo  {journal} {Astronomy Education Review}\ }\textbf {\bibinfo
  {volume} {6}},\ \bibinfo {pages} {133} (\bibinfo {year} {2007})}\BibitemShut
  {NoStop}%
\bibitem [{\citenamefont {Bailey}\ \emph
  {et~al.}(2012{\natexlab{a}})\citenamefont {Bailey}, \citenamefont {Johnson},
  \citenamefont {Prather},\ and\ \citenamefont {Slater}}]{Bailey2012}%
  \BibitemOpen
  \bibfield  {author} {\bibinfo {author} {\bibfnamefont {J.~M.}\ \bibnamefont
  {Bailey}}, \bibinfo {author} {\bibfnamefont {B.}~\bibnamefont {Johnson}},
  \bibinfo {author} {\bibfnamefont {E.~E.}\ \bibnamefont {Prather}}, \ and\
  \bibinfo {author} {\bibfnamefont {T.~F.}\ \bibnamefont {Slater}},\ }\href
  {\doibase 10.1080/09500693.2011.589869} {\bibfield  {journal} {\bibinfo
  {journal} {International Journal of Science Education}\ }\textbf {\bibinfo
  {volume} {34}},\ \bibinfo {pages} {2257} (\bibinfo {year}
  {2012}{\natexlab{a}})},\ \Eprint
  {http://arxiv.org/abs/http://dx.doi.org/10.1080/09500693.2011.589869}
  {http://dx.doi.org/10.1080/09500693.2011.589869} \BibitemShut {NoStop}%
\bibitem [{\citenamefont {{Lopresto}}\ and\ \citenamefont
  {{Murrell}}(2009)}]{Lopresto:2009}%
  \BibitemOpen
  \bibfield  {author} {\bibinfo {author} {\bibfnamefont {M.~C.}\ \bibnamefont
  {{Lopresto}}}\ and\ \bibinfo {author} {\bibfnamefont {S.~R.}\ \bibnamefont
  {{Murrell}}},\ }\href {\doibase 10.3847/AER2009014} {\bibfield  {journal}
  {\bibinfo  {journal} {Astronomy Education Review}\ }\textbf {\bibinfo
  {volume} {8}},\ \bibinfo {pages} {010105} (\bibinfo {year}
  {2009})}\BibitemShut {NoStop}%
\bibitem [{\citenamefont {{Lindell}}(2001)}]{Lindell:2001}%
  \BibitemOpen
  \bibfield  {author} {\bibinfo {author} {\bibfnamefont {R.~S.}\ \bibnamefont
  {{Lindell}}},\ }\emph {\bibinfo {title} {{Enhancing college students'
  understanding of lunar phases}}},\ \href@noop {} {Ph.D. thesis},\ \bibinfo
  {school} {The University of Nebraska - Lincoln} (\bibinfo {year}
  {2001})\BibitemShut {NoStop}%
\bibitem [{\citenamefont {{Lindell}}\ and\ \citenamefont
  {{Sommer}}(2004)}]{Lindell:2004}%
  \BibitemOpen
  \bibfield  {author} {\bibinfo {author} {\bibfnamefont {R.~S.}\ \bibnamefont
  {{Lindell}}}\ and\ \bibinfo {author} {\bibfnamefont {S.~R.}\ \bibnamefont
  {{Sommer}}},\ }in\ \href {\doibase 10.1063/1.1807257} {\emph {\bibinfo
  {booktitle} {American Institute of Physics Conference Series}}},\ \bibinfo
  {series} {American Institute of Physics Conference Series}, Vol.\ \bibinfo
  {volume} {720},\ \bibinfo {editor} {edited by\ \bibinfo {editor}
  {\bibfnamefont {J.}~\bibnamefont {{Marx}}}, \bibinfo {editor} {\bibfnamefont
  {K.}~\bibnamefont {{Cummings}}}, \ and\ \bibinfo {editor} {\bibfnamefont
  {S.}~\bibnamefont {{Franklin}}}}\ (\bibinfo {year} {2004})\ pp.\ \bibinfo
  {pages} {73--76}\BibitemShut {NoStop}%
\bibitem [{\citenamefont {{Williamson}}\ \emph {et~al.}(2013)\citenamefont
  {{Williamson}}, \citenamefont {{Willoughby}},\ and\ \citenamefont
  {{Prather}}}]{2013AEdRv..12a0107W}%
  \BibitemOpen
  \bibfield  {author} {\bibinfo {author} {\bibfnamefont {K.~E.}\ \bibnamefont
  {{Williamson}}}, \bibinfo {author} {\bibfnamefont {S.}~\bibnamefont
  {{Willoughby}}}, \ and\ \bibinfo {author} {\bibfnamefont {E.~E.}\
  \bibnamefont {{Prather}}},\ }\href {\doibase 10.3847/AER2012045} {\bibfield
  {journal} {\bibinfo  {journal} {Astronomy Education Review}\ }\textbf
  {\bibinfo {volume} {12}},\ \bibinfo {pages} {010107} (\bibinfo {year}
  {2013})}\BibitemShut {NoStop}%
\bibitem [{\citenamefont {Williamson}\ and\ \citenamefont
  {Willoughby}(2012)}]{Williamson:2012}%
  \BibitemOpen
  \bibfield  {author} {\bibinfo {author} {\bibfnamefont {K.~E.}\ \bibnamefont
  {Williamson}}\ and\ \bibinfo {author} {\bibfnamefont {S.}~\bibnamefont
  {Willoughby}},\ }\href {\doibase http://dx.doi.org/10.3847/AER2011025}
  {\bibfield  {journal} {\bibinfo  {journal} {Astronomy Education Review}\
  }\textbf {\bibinfo {volume} {11}},\ \bibinfo {eid} {010105} (\bibinfo {year}
  {2012})}\BibitemShut {NoStop}%
\bibitem [{\citenamefont {{Wallace}}\ \emph
  {et~al.}(2011{\natexlab{a}})\citenamefont {{Wallace}}, \citenamefont
  {{Prather}},\ and\ \citenamefont {{Duncan}}}]{2011AEdRv..10a0106W}%
  \BibitemOpen
  \bibfield  {author} {\bibinfo {author} {\bibfnamefont {C.~S.}\ \bibnamefont
  {{Wallace}}}, \bibinfo {author} {\bibfnamefont {E.~E.}\ \bibnamefont
  {{Prather}}}, \ and\ \bibinfo {author} {\bibfnamefont {D.~K.}\ \bibnamefont
  {{Duncan}}},\ }\href {\doibase 10.3847/AER2011029} {\bibfield  {journal}
  {\bibinfo  {journal} {Astronomy Education Review}\ }\textbf {\bibinfo
  {volume} {10}},\ \bibinfo {pages} {010106} (\bibinfo {year}
  {2011}{\natexlab{a}})}\BibitemShut {NoStop}%
\bibitem [{\citenamefont {{Wallace}}\ \emph
  {et~al.}(2011{\natexlab{b}})\citenamefont {{Wallace}}, \citenamefont
  {{Prather}},\ and\ \citenamefont {{Duncan}}}]{2011AEdRv..10a0107W}%
  \BibitemOpen
  \bibfield  {author} {\bibinfo {author} {\bibfnamefont {C.~S.}\ \bibnamefont
  {{Wallace}}}, \bibinfo {author} {\bibfnamefont {E.~E.}\ \bibnamefont
  {{Prather}}}, \ and\ \bibinfo {author} {\bibfnamefont {D.~K.}\ \bibnamefont
  {{Duncan}}},\ }\href {\doibase 10.3847/AER2011030} {\bibfield  {journal}
  {\bibinfo  {journal} {Astronomy Education Review}\ }\textbf {\bibinfo
  {volume} {10}},\ \bibinfo {pages} {010107} (\bibinfo {year}
  {2011}{\natexlab{b}})}\BibitemShut {NoStop}%
\bibitem [{\citenamefont {{Wallace}}\ \emph
  {et~al.}(2012{\natexlab{a}})\citenamefont {{Wallace}}, \citenamefont
  {{Prather}},\ and\ \citenamefont {{Duncan}}}]{2012AEdRv..11a0103W}%
  \BibitemOpen
  \bibfield  {author} {\bibinfo {author} {\bibfnamefont {C.~S.}\ \bibnamefont
  {{Wallace}}}, \bibinfo {author} {\bibfnamefont {E.~E.}\ \bibnamefont
  {{Prather}}}, \ and\ \bibinfo {author} {\bibfnamefont {D.~K.}\ \bibnamefont
  {{Duncan}}},\ }\href {\doibase 10.3847/AER2011031} {\bibfield  {journal}
  {\bibinfo  {journal} {Astronomy Education Review}\ }\textbf {\bibinfo
  {volume} {11}},\ \bibinfo {pages} {010103} (\bibinfo {year}
  {2012}{\natexlab{a}})}\BibitemShut {NoStop}%
\bibitem [{\citenamefont {{Wallace}}\ \emph
  {et~al.}(2012{\natexlab{b}})\citenamefont {{Wallace}}, \citenamefont
  {{Prather}},\ and\ \citenamefont {{Duncan}}}]{2012AEdRv..11a0104W}%
  \BibitemOpen
  \bibfield  {author} {\bibinfo {author} {\bibfnamefont {C.~S.}\ \bibnamefont
  {{Wallace}}}, \bibinfo {author} {\bibfnamefont {E.~E.}\ \bibnamefont
  {{Prather}}}, \ and\ \bibinfo {author} {\bibfnamefont {D.~K.}\ \bibnamefont
  {{Duncan}}},\ }\href {\doibase 10.3847/AER2011032} {\bibfield  {journal}
  {\bibinfo  {journal} {Astronomy Education Review}\ }\textbf {\bibinfo
  {volume} {11}},\ \bibinfo {pages} {010104} (\bibinfo {year}
  {2012}{\natexlab{b}})}\BibitemShut {NoStop}%
\bibitem [{\citenamefont {{Prather}}\ \emph
  {et~al.}(2002{\natexlab{a}})\citenamefont {{Prather}}, \citenamefont
  {{Slater}},\ and\ \citenamefont {{Offerdahl}}}]{2002AEdRv...1b..28P}%
  \BibitemOpen
  \bibfield  {author} {\bibinfo {author} {\bibfnamefont {E.~E.}\ \bibnamefont
  {{Prather}}}, \bibinfo {author} {\bibfnamefont {T.~F.}\ \bibnamefont
  {{Slater}}}, \ and\ \bibinfo {author} {\bibfnamefont {E.~G.}\ \bibnamefont
  {{Offerdahl}}},\ }\href@noop {} {\bibfield  {journal} {\bibinfo  {journal}
  {Astronomy Education Review}\ }\textbf {\bibinfo {volume} {1}},\ \bibinfo
  {pages} {020000} (\bibinfo {year} {2002}{\natexlab{a}})}\BibitemShut
  {NoStop}%
\bibitem [{\citenamefont {{Trumper}}(2001)}]{Trumper:2001b}%
  \BibitemOpen
  \bibfield  {author} {\bibinfo {author} {\bibfnamefont {R.}~\bibnamefont
  {{Trumper}}},\ }\href@noop {} {\bibfield  {journal} {\bibinfo  {journal}
  {Journal of Science Education and Technology}\ }\textbf {\bibinfo {volume}
  {10}},\ \bibinfo {pages} {189} (\bibinfo {year} {2001})}\BibitemShut
  {NoStop}%
\bibitem [{\citenamefont {{Waller}}\ and\ \citenamefont
  {{Slater}}(2011)}]{Waller:2011}%
  \BibitemOpen
  \bibfield  {author} {\bibinfo {author} {\bibfnamefont {W.~H.}\ \bibnamefont
  {{Waller}}}\ and\ \bibinfo {author} {\bibfnamefont {T.~F.}\ \bibnamefont
  {{Slater}}},\ }\href {\doibase 10.5408/1.3651408} {\bibfield  {journal}
  {\bibinfo  {journal} {Journal of Geoscience Education}\ }\textbf {\bibinfo
  {volume} {59}},\ \bibinfo {pages} {176} (\bibinfo {year} {2011})}\BibitemShut
  {NoStop}%
\bibitem [{\citenamefont {{Sugarman}}\ \emph {et~al.}(2011)\citenamefont
  {{Sugarman}}, \citenamefont {{Impey}}, \citenamefont {{Buxner}},\ and\
  \citenamefont {{Antonellis}}}]{2011AEdRv..10a0101S}%
  \BibitemOpen
  \bibfield  {author} {\bibinfo {author} {\bibfnamefont {H.}~\bibnamefont
  {{Sugarman}}}, \bibinfo {author} {\bibfnamefont {C.}~\bibnamefont {{Impey}}},
  \bibinfo {author} {\bibfnamefont {S.}~\bibnamefont {{Buxner}}}, \ and\
  \bibinfo {author} {\bibfnamefont {J.}~\bibnamefont {{Antonellis}}},\ }\href
  {\doibase 10.3847/AER2010040} {\bibfield  {journal} {\bibinfo  {journal}
  {Astronomy Education Review}\ }\textbf {\bibinfo {volume} {10}},\ \bibinfo
  {pages} {010101} (\bibinfo {year} {2011})}\BibitemShut {NoStop}%
\bibitem [{\citenamefont {Caton}(2013)}]{Caton:2013}%
  \BibitemOpen
  \bibfield  {author} {\bibinfo {author} {\bibfnamefont {D.}~\bibnamefont
  {Caton}},\ }\href {\doibase http://dx.doi.org/10.1119/1.4824939} {\bibfield
  {journal} {\bibinfo  {journal} {The Physics Teacher}\ }\textbf {\bibinfo
  {volume} {51}},\ \bibinfo {pages} {470} (\bibinfo {year} {2013})}\BibitemShut
  {NoStop}%
\bibitem [{\citenamefont {Volkwyn}(2005)}]{Volkwyn:2005}%
  \BibitemOpen
  \bibfield  {author} {\bibinfo {author} {\bibfnamefont {T.~S.}\ \bibnamefont
  {Volkwyn}},\ }\emph {\bibinfo {title} {First year students' understanding of
  measurement in physics laboratory work}},\ \href@noop {} {Master's thesis},\
  \bibinfo  {school} {University of Cape Town} (\bibinfo {year}
  {2005})\BibitemShut {NoStop}%
\bibitem [{\citenamefont {White}\ \emph {et~al.}(1999)\citenamefont {White},
  \citenamefont {Elby}, \citenamefont {Frederiksen},\ and\ \citenamefont
  {Schwarz}}]{White:1999}%
  \BibitemOpen
  \bibfield  {author} {\bibinfo {author} {\bibfnamefont {B.}~\bibnamefont
  {White}}, \bibinfo {author} {\bibfnamefont {A.}~\bibnamefont {Elby}},
  \bibinfo {author} {\bibfnamefont {J.}~\bibnamefont {Frederiksen}}, \ and\
  \bibinfo {author} {\bibfnamefont {C.}~\bibnamefont {Schwarz}},\ }\href@noop
  {} {\enquote {\bibinfo {title} {The epistemological beliefs assessment for
  physical science},}\ } (\bibinfo {year} {1999}),\ \bibinfo {note} {presented
  at the American Education Research Association, Montreal
  (unpublished)}\BibitemShut {NoStop}%
\bibitem [{\citenamefont {{Elby}}(2001)}]{Elby:2001}%
  \BibitemOpen
  \bibfield  {author} {\bibinfo {author} {\bibfnamefont {A.}~\bibnamefont
  {{Elby}}},\ }\href {\doibase 10.1119/1.1377283} {\bibfield  {journal}
  {\bibinfo  {journal} {American Journal of Physics}\ }\textbf {\bibinfo
  {volume} {69}},\ \bibinfo {pages} {54} (\bibinfo {year} {2001})}\BibitemShut
  {NoStop}%
\bibitem [{\citenamefont {Nwosu}(2012)}]{Nwosu:2012}%
  \BibitemOpen
  \bibfield  {author} {\bibinfo {author} {\bibfnamefont {V.}~\bibnamefont
  {Nwosu}},\ }\emph {\bibinfo {title} {A study of postgraduate students in an
  astrophysics bridging year: identifying contradictions in a complex
  system}},\ \href@noop {} {Ph.D. thesis},\ \bibinfo  {school} {University of
  Cape Town} (\bibinfo {year} {2012})\BibitemShut {NoStop}%
\bibitem [{\citenamefont {{Allie}}\ \emph {et~al.}(2008)\citenamefont
  {{Allie}}, \citenamefont {{Demaree}}, \citenamefont {{Taylor}}, \citenamefont
  {{Lubben}},\ and\ \citenamefont {{Buffler}}}]{Allie:2008}%
  \BibitemOpen
  \bibfield  {author} {\bibinfo {author} {\bibfnamefont {S.}~\bibnamefont
  {{Allie}}}, \bibinfo {author} {\bibfnamefont {D.}~\bibnamefont {{Demaree}}},
  \bibinfo {author} {\bibfnamefont {J.}~\bibnamefont {{Taylor}}}, \bibinfo
  {author} {\bibfnamefont {F.}~\bibnamefont {{Lubben}}}, \ and\ \bibinfo
  {author} {\bibfnamefont {A.}~\bibnamefont {{Buffler}}},\ }\href {\doibase
  10.1063/1.3021268} {\bibfield  {journal} {\bibinfo  {journal} {AIP Conference
  Proceedings}\ }\textbf {\bibinfo {volume} {1064}},\ \bibinfo {pages} {3}
  (\bibinfo {year} {2008})}\BibitemShut {NoStop}%
\bibitem [{\citenamefont {Allie}\ and\ \citenamefont
  {Demaree}(2010)}]{Allie:2010}%
  \BibitemOpen
  \bibfield  {author} {\bibinfo {author} {\bibfnamefont {S.}~\bibnamefont
  {Allie}}\ and\ \bibinfo {author} {\bibfnamefont {D.}~\bibnamefont
  {Demaree}},\ }\href {\doibase 10.1063/1.3515198} {\bibfield  {journal}
  {\bibinfo  {journal} {AIP Conference Proceedings}\ }\textbf {\bibinfo
  {volume} {1289}},\ \bibinfo {pages} {1} (\bibinfo {year} {2010})}\BibitemShut
  {NoStop}%
\bibitem [{\citenamefont {Johnson}(1964)}]{Johnson:1964}%
  \BibitemOpen
  \bibfield  {author} {\bibinfo {author} {\bibfnamefont {P.~E.}\ \bibnamefont
  {Johnson}},\ }\href {\doibase 10.1037/h0042666} {\bibfield  {journal}
  {\bibinfo  {journal} {Journal of Educational Psychology}\ }\textbf {\bibinfo
  {volume} {55}},\ \bibinfo {pages} {84} (\bibinfo {year} {1964})}\BibitemShut
  {NoStop}%
\bibitem [{\citenamefont {Nelson}\ \emph {et~al.}(2000)\citenamefont {Nelson},
  \citenamefont {McEvoy},\ and\ \citenamefont {Dennis}}]{Nelson:2000}%
  \BibitemOpen
  \bibfield  {author} {\bibinfo {author} {\bibfnamefont {D.~L.}\ \bibnamefont
  {Nelson}}, \bibinfo {author} {\bibfnamefont {C.~L.}\ \bibnamefont {McEvoy}},
  \ and\ \bibinfo {author} {\bibfnamefont {S.}~\bibnamefont {Dennis}},\ }\href
  {\doibase 10.3758/BF03209337} {\bibfield  {journal} {\bibinfo  {journal}
  {Memory \& Cognition}\ }\textbf {\bibinfo {volume} {28}},\ \bibinfo {pages}
  {887} (\bibinfo {year} {2000})}\BibitemShut {NoStop}%
\bibitem [{Note5()}]{Note5}%
  \BibitemOpen
  \bibinfo {note} {This served to make students accountable for their answers
  and not to give whimsical answers; from the responses, it appeared that the
  instrument was indeed answered seriously.}\BibitemShut {Stop}%
\bibitem [{\citenamefont {Corbin}\ and\ \citenamefont
  {Strauss}(1990)}]{Corbin:1990}%
  \BibitemOpen
  \bibfield  {author} {\bibinfo {author} {\bibnamefont {Corbin}}\ and\ \bibinfo
  {author} {\bibfnamefont {A.}~\bibnamefont {Strauss}},\ }\href@noop {}
  {\bibfield  {journal} {\bibinfo  {journal} {Qualitative Sociology}\ ,\
  \bibinfo {pages} {3}} (\bibinfo {year} {1990})}\BibitemShut {NoStop}%
\bibitem [{\citenamefont {Strauss}\ and\ \citenamefont
  {Corbin}(1997)}]{Strauss:1997}%
  \BibitemOpen
  \bibfield  {author} {\bibinfo {author} {\bibfnamefont {A.~L.}\ \bibnamefont
  {Strauss}}\ and\ \bibinfo {author} {\bibfnamefont {J.~M.}\ \bibnamefont
  {Corbin}},\ }\href@noop {} {\emph {\bibinfo {title} {Grounded theory in
  practice}}},\ \bibinfo {edition} {2nd}\ ed.\ (\bibinfo  {publisher} {Thousand
  Oaks, CA, USA},\ \bibinfo {year} {1997})\BibitemShut {NoStop}%
\bibitem [{Note6()}]{Note6}%
  \BibitemOpen
  \bibinfo {note} {For example, whereas words such as ``stars'', ``galaxies'',
  and ``planets'' all saw drops in the proportion of students citing them,
  these words were very popular ($>50\%$) both pre- and post-course, and indeed
  their ranks remained essentially unchanged.}\BibitemShut {Stop}%
\bibitem [{\citenamefont {Trouille}\ \emph {et~al.}(2013)\citenamefont
  {Trouille}, \citenamefont {Coble}, \citenamefont {Cochran}, \citenamefont
  {Bailey}, \citenamefont {Camarillo}, \citenamefont {Nickerson},\ and\
  \citenamefont {Cominsky}}]{Trouille:2013}%
  \BibitemOpen
  \bibfield  {author} {\bibinfo {author} {\bibfnamefont {L.~E.}\ \bibnamefont
  {Trouille}}, \bibinfo {author} {\bibfnamefont {K.}~\bibnamefont {Coble}},
  \bibinfo {author} {\bibfnamefont {G.~L.}\ \bibnamefont {Cochran}}, \bibinfo
  {author} {\bibfnamefont {J.~M.}\ \bibnamefont {Bailey}}, \bibinfo {author}
  {\bibfnamefont {C.~T.}\ \bibnamefont {Camarillo}}, \bibinfo {author}
  {\bibfnamefont {M.~D.}\ \bibnamefont {Nickerson}}, \ and\ \bibinfo {author}
  {\bibfnamefont {L.~R.}\ \bibnamefont {Cominsky}},\ }\href {\doibase
  http://dx.doi.org/10.3847/AER2013016} {\bibfield  {journal} {\bibinfo
  {journal} {Astronomy Education Review}\ }\textbf {\bibinfo {volume} {12}},\
  \bibinfo {eid} {010110} (\bibinfo {year} {2013})}\BibitemShut {NoStop}%
\bibitem [{\citenamefont {{Prather}}\ \emph
  {et~al.}(2002{\natexlab{b}})\citenamefont {{Prather}}, \citenamefont
  {{Slater}},\ and\ \citenamefont {{Offerdahl}}}]{Prather:2002}%
  \BibitemOpen
  \bibfield  {author} {\bibinfo {author} {\bibfnamefont {E.~E.}\ \bibnamefont
  {{Prather}}}, \bibinfo {author} {\bibfnamefont {T.~F.}\ \bibnamefont
  {{Slater}}}, \ and\ \bibinfo {author} {\bibfnamefont {E.~G.}\ \bibnamefont
  {{Offerdahl}}},\ }\href@noop {} {\bibfield  {journal} {\bibinfo  {journal}
  {Astronomy Education Review}\ }\textbf {\bibinfo {volume} {1}},\ \bibinfo
  {pages} {020000} (\bibinfo {year} {2002}{\natexlab{b}})}\BibitemShut
  {NoStop}%
\bibitem [{\citenamefont {Noce}\ \emph {et~al.}(1988)\citenamefont {Noce},
  \citenamefont {Torosantucci},\ and\ \citenamefont {Vicentini}}]{Noce:1988}%
  \BibitemOpen
  \bibfield  {author} {\bibinfo {author} {\bibfnamefont {G.}~\bibnamefont
  {Noce}}, \bibinfo {author} {\bibfnamefont {G.}~\bibnamefont {Torosantucci}},
  \ and\ \bibinfo {author} {\bibfnamefont {M.}~\bibnamefont {Vicentini}},\
  }\href {\doibase 10.1080/0950069880100106} {\bibfield  {journal} {\bibinfo
  {journal} {International Journal of Science Education}\ }\textbf {\bibinfo
  {volume} {10}},\ \bibinfo {pages} {61} (\bibinfo {year} {1988})}\BibitemShut
  {NoStop}%
\bibitem [{\citenamefont {Galili}\ and\ \citenamefont
  {Kaplan}(1996)}]{Galili:1996}%
  \BibitemOpen
  \bibfield  {author} {\bibinfo {author} {\bibfnamefont {I.}~\bibnamefont
  {Galili}}\ and\ \bibinfo {author} {\bibfnamefont {D.}~\bibnamefont
  {Kaplan}},\ }\href {\doibase
  10.1002/(SICI)1098-237X(199607)80:4<457::AID-SCE5>3.0.CO;2-C} {\bibfield
  {journal} {\bibinfo  {journal} {Science Education}\ }\textbf {\bibinfo
  {volume} {80}},\ \bibinfo {pages} {457} (\bibinfo {year} {1996})}\BibitemShut
  {NoStop}%
\bibitem [{\citenamefont {Bardar}(2007)}]{Bardar:2007}%
  \BibitemOpen
  \bibfield  {author} {\bibinfo {author} {\bibfnamefont {E.~M.}\ \bibnamefont
  {Bardar}},\ }\href {\doibase http://dx.doi.org/10.3847/AER2007019} {\bibfield
   {journal} {\bibinfo  {journal} {Astronomy Education Review}\ }\textbf
  {\bibinfo {volume} {6}},\ \bibinfo {pages} {75} (\bibinfo {year}
  {2007})}\BibitemShut {NoStop}%
\bibitem [{\citenamefont {Plait}(2002)}]{plait:2002}%
  \BibitemOpen
  \bibfield  {author} {\bibinfo {author} {\bibfnamefont {P.~C.}\ \bibnamefont
  {Plait}},\ }\href@noop {} {\emph {\bibinfo {title} {Bad astronomy:
  misconceptions and misuses revealed, from astrology to the moon landing
  ``hoax''}}}\ (\bibinfo  {publisher} {John Wiley \& Sons},\ \bibinfo {year}
  {2002})\BibitemShut {NoStop}%
\bibitem [{\citenamefont {Marx}\ and\ \citenamefont
  {Lindell}(2004)}]{marx:2004}%
  \BibitemOpen
  \bibfield  {author} {\bibinfo {author} {\bibfnamefont {J.}~\bibnamefont
  {Marx}}\ and\ \bibinfo {author} {\bibfnamefont {R.}~\bibnamefont {Lindell}},\
  }\href {\doibase http://dx.doi.org/10.1063/1.1807256} {\bibfield  {journal}
  {\bibinfo  {journal} {AIP Conference Proceedings}\ }\textbf {\bibinfo
  {volume} {720}},\ \bibinfo {pages} {69} (\bibinfo {year} {2004})}\BibitemShut
  {NoStop}%
\bibitem [{\citenamefont {Coble}\ \emph {et~al.}(2013)\citenamefont {Coble},
  \citenamefont {Camarillo}, \citenamefont {Trouille}, \citenamefont {Bailey},
  \citenamefont {Cochran}, \citenamefont {Nickerson},\ and\ \citenamefont
  {Cominsky}}]{Coble:2013}%
  \BibitemOpen
  \bibfield  {author} {\bibinfo {author} {\bibfnamefont {K.}~\bibnamefont
  {Coble}}, \bibinfo {author} {\bibfnamefont {C.~T.}\ \bibnamefont
  {Camarillo}}, \bibinfo {author} {\bibfnamefont {L.~E.}\ \bibnamefont
  {Trouille}}, \bibinfo {author} {\bibfnamefont {J.~M.}\ \bibnamefont
  {Bailey}}, \bibinfo {author} {\bibfnamefont {G.~L.}\ \bibnamefont {Cochran}},
  \bibinfo {author} {\bibfnamefont {M.~D.}\ \bibnamefont {Nickerson}}, \ and\
  \bibinfo {author} {\bibfnamefont {L.~R.}\ \bibnamefont {Cominsky}},\ }\href
  {\doibase http://dx.doi.org/10.3847/AER2012038} {\bibfield  {journal}
  {\bibinfo  {journal} {Astronomy Education Review}\ }\textbf {\bibinfo
  {volume} {12}},\ \bibinfo {eid} {010102} (\bibinfo {year}
  {2013})}\BibitemShut {NoStop}%
\bibitem [{\citenamefont {Bailey}\ \emph
  {et~al.}(2012{\natexlab{b}})\citenamefont {Bailey}, \citenamefont {Coble},
  \citenamefont {Cochran}, \citenamefont {Larrieu}, \citenamefont {Sanchez},\
  and\ \citenamefont {Cominsky}}]{Bailey:2012b}%
  \BibitemOpen
  \bibfield  {author} {\bibinfo {author} {\bibfnamefont {J.~M.}\ \bibnamefont
  {Bailey}}, \bibinfo {author} {\bibfnamefont {K.}~\bibnamefont {Coble}},
  \bibinfo {author} {\bibfnamefont {G.}~\bibnamefont {Cochran}}, \bibinfo
  {author} {\bibfnamefont {D.}~\bibnamefont {Larrieu}}, \bibinfo {author}
  {\bibfnamefont {R.}~\bibnamefont {Sanchez}}, \ and\ \bibinfo {author}
  {\bibfnamefont {L.~R.}\ \bibnamefont {Cominsky}},\ }\href {\doibase
  http://dx.doi.org/10.3847/AER2012029} {\bibfield  {journal} {\bibinfo
  {journal} {Astronomy Education Review}\ }\textbf {\bibinfo {volume} {11}},\
  \bibinfo {eid} {010302} (\bibinfo {year} {2012}{\natexlab{b}})}\BibitemShut
  {NoStop}%
\bibitem [{Note7()}]{Note7}%
  \BibitemOpen
  \bibinfo {note} {The null hypothesis was $r=0$, with alternative hypothesis
  $r\not =0$.}\BibitemShut {Stop}%
\bibitem [{\citenamefont {{Matthews}}(2009)}]{Matthews:2009}%
  \BibitemOpen
  \bibfield  {author} {\bibinfo {author} {\bibfnamefont {M.~R.}\ \bibnamefont
  {{Matthews}}},\ }\href {\doibase 10.1007/s11191-008-9170-6} {\bibfield
  {journal} {\bibinfo  {journal} {Science \& Education}\ }\textbf {\bibinfo
  {volume} {18}},\ \bibinfo {pages} {641} (\bibinfo {year} {2009})}\BibitemShut
  {NoStop}%
\bibitem [{\citenamefont {Perry}(1998)}]{Perry:1998}%
  \BibitemOpen
  \bibfield  {author} {\bibinfo {author} {\bibfnamefont {W.~G.}\ \bibnamefont
  {Perry}},\ }\href@noop {} {\emph {\bibinfo {title} {Forms of ethical and
  intellectual development in the college years: a scheme}}},\ \bibinfo
  {edition} {1st}\ ed.\ (\bibinfo  {publisher} {Jossey-Bass},\ \bibinfo {year}
  {1998})\BibitemShut {NoStop}%
\bibitem [{\citenamefont {{Finster}}(1989)}]{Finster:1989}%
  \BibitemOpen
  \bibfield  {author} {\bibinfo {author} {\bibfnamefont {D.~C.}\ \bibnamefont
  {{Finster}}},\ }\href {\doibase 10.1021/ed066p659} {\bibfield  {journal}
  {\bibinfo  {journal} {Journal of Chemical Education}\ }\textbf {\bibinfo
  {volume} {66}},\ \bibinfo {pages} {659} (\bibinfo {year} {1989})}\BibitemShut
  {NoStop}%
\bibitem [{\citenamefont {{Finster}}(1991)}]{Finster:1991}%
  \BibitemOpen
  \bibfield  {author} {\bibinfo {author} {\bibfnamefont {D.~C.}\ \bibnamefont
  {{Finster}}},\ }\href {\doibase 10.1021/ed068p752} {\bibfield  {journal}
  {\bibinfo  {journal} {Journal of Chemical Education}\ }\textbf {\bibinfo
  {volume} {68}},\ \bibinfo {pages} {752} (\bibinfo {year} {1991})}\BibitemShut
  {NoStop}%
\bibitem [{\citenamefont {Kagee}\ \emph {et~al.}(2010)\citenamefont {Kagee},
  \citenamefont {Allie},\ and\ \citenamefont {Lesch}}]{Kagee:2010}%
  \BibitemOpen
  \bibfield  {author} {\bibinfo {author} {\bibfnamefont {A.}~\bibnamefont
  {Kagee}}, \bibinfo {author} {\bibfnamefont {S.}~\bibnamefont {Allie}}, \ and\
  \bibinfo {author} {\bibfnamefont {A.}~\bibnamefont {Lesch}},\ }\href@noop {}
  {\bibfield  {journal} {\bibinfo  {journal} {South African Journal of
  Psychology}\ }\textbf {\bibinfo {volume} {40}},\ \bibinfo {pages} {272}
  (\bibinfo {year} {2010})}\BibitemShut {NoStop}%
\end{thebibliography}%
%--------------------------------------------------------------------------------------------------
\end{document}